\documentclass[letter,11pt]{article}
\usepackage{jinsttemplate}

\usepackage{graphicx}
\usepackage{caption}
\usepackage{subcaption}
\usepackage{float}
\usepackage{booktabs} 
\usepackage{enumitem}
\usepackage[skip=10pt]{caption}
\usepackage{url}
\DeclareUnicodeCharacter{223C}{$\sim$}
\usepackage{verbatim}
\title{Production and installation of wavelength-shifting reflective light enhancers for the Short-Baseline Near Detector}
\collaboration{SBND Collaboration}

\author[14]{R.~Acciarri}
\author[39]{L.~Aliaga-Soplin}
\author[8]{R.~Alvarez-Garrote}
\author[17]{D.~Andrade Aldana}
\author[22]{C.~Andreopoulos}
\author[4]{A.~Antonakis}
\author[28]{S.~Balasubramanian}
\author[30]{A.~Barnard}
\author[14]{V.~Basque}
\author[25,18]{J.~Bateman}

% ...
\affiliation[4]{University of California, Santa Barbara CA, 93106, USA}
% ...
\affiliation[8]{CIEMAT, Centro de Investigaciones Energ\'{e}ticas, Medioambientales y Tecnol\'{o}gicas, Madrid E-28040, Spain}
% ...
\affiliation[14]{Fermi National Accelerator Laboratory, Batavia, IL 60510, USA}
% ...
\affiliation[17]{Illinois Institute of Technology, Chicago, IL 60616, USA}
\affiliation[18]{Imperial College London, London SW7 2AZ, United Kingdom}
% ...
\affiliation[22]{University of Liverpool, Liverpool L69 7ZE, United Kingdom}
% ...
\affiliation[25]{University of Manchester, Manchester M13 9PL, United Kingdom}
% ...
\affiliation[28]{Mount Holyoke College, South Hadley, MA 01075, USA}
% ...
\affiliation[30]{University of Oxford, Oxford OX1 3RH, United Kingdom}
% ...
\affiliation[39]{University of Texas at Arlington, TX 76019, USA}
% ...

\abstract{We report on the design, production, and installation of a wavelength-shifting reflective system on the cathode of the Short-Baseline Near Detector (SBND), a liquid argon time projection chamber located along the Fermilab Booster Neutrino Beam. To increase and homogenize scintillation-light collection, 64 double-sided plates were fabricated from FR4, laminated with specular reflector film and coated with ∼300~$\upmu$g/cm² of tetraphenyl butadiene (TPB) wavelength shifter using controlled physical vapor deposition. The coating uniformity was validated through dedicated measurements of deposited mass and profilometry studies. Because exposure to ambient blue/UV light could degrade the TPB, protective filtering and controlled storage conditions were implemented during handling and installation. The coated plates were assembled between conductive meshes for high-voltage compatibility and installed in situ during detector integration. This system constitutes the largest TPB-coated area deployed in a neutrino detector. It operates in conjunction with SBND’s photon detection system, which consists of photomultiplier tubes and X-ARAPUCAs. Early light-collection measurements show high uniformity and light response across the detector, supporting improved triggering, calorimetry, and position reconstruction in SBND.}

\emailAdd{sbnd\_info@fnal.gov}


\begin{document}

\maketitle

\section{Introduction}

Scintillation light produced by particle interactions in liquid argon time projection chambers (LArTPCs) enables fast timing, powerful triggering and background-rejection capabilities, as well as light-only calorimetry.~\cite{MicroBooNE:2023ldj, SBND:2024vgn, Boulay:2019eal, Fields:2020wge}. Additionally, combining scintillation light with the ionization charge produced by charged-particle interactions has been shown to improve energy resolution~\cite{LArIAT:2019gdz}, which is particularly advantageous for studies of low-energy interactions. However, photon detection systems (PDS) in most existing LArTPCs are limited by their photosensor coverage, restricting their effectiveness, particularly for calorimetric applications. Ongoing efforts across noble-element detectors, including single- and dual-phase detectors, focus on developing technologies to increase the amount of scintillation light collected. These efforts are an important aspect of R\&D for future liquid-noble detectors, such as the far-detector modules of the Deep Underground Neutrino Experiment (DUNE)~\cite{DUNE:2015lol}. The Short-Baseline Near Detector (SBND), a LArTPC and a part of the Short-Baseline Neutrino (SBN) Program at Fermilab~\cite{SBND}, has installed an entire wall of wavelength-shifting reflective plates within the cathode of its LArTPC to achieve this goal. These passive elements enhance the performance of the PDS by increasing light collection through wavelength-shifting and by reflecting photons that would otherwise be absorbed by the cathode, thereby enabling more uniform light collection. This article reports on the design, production, assembly, and installation of the wavelength-shifting reflective plates in SBND, as well as the motivation for their inclusion. 

\section{The SBND experiment}

The SBND experiment is a LArTPC located 110~m downstream of the Booster Neutrino Beam (BNB) target at Fermilab \cite{sbnd_detector_paper}. As the near detector of the SBN Program, SBND measures the neutrino fluxes close to the production source, enabling precision studies of short-baseline neutrino oscillations in conjunction with the far detector, ICARUS~\cite{ICARUS:2023gpo}. SBND is recording an unprecedented number of neutrino-induced interactions within its active volume, which also facilitates high-statistics measurements of neutrino-argon cross sections and competitive searches for physics beyond the Standard Model.

The TPC of 4~m height by 5~m length by 4~m width contains 112 tonnes of active volume of liquid argon. It began its data collection campaign in 2024. The SBND LArTPC consists of two independent drift volumes joined at a central cathode, which contains 16 windows designed to be covered with stainless steel meshes. Each drift volume is terminated on the opposite side by an anode plane comprising three wire planes, which are used to collect the ionization charge produced by charged particle interactions in the liquid argon. A uniform electric field of 500~V/cm is maintained between the cathode and each anode by keeping the cathode at a potential of -100 kV, allowing the drift of ionization electrons. The field uniformity is established and preserved by a surrounding field cage. The LArTPC is housed inside a membrane cryostat, with a dedicated cryogenic system that maintains the liquid argon at its nominal operating temperature of 87~K. The cryogenic system also includes a series of purification filters that continuously remove electronegative contaminants, ensuring high argon purity during sustained detector operation.

\subsection{Photon detection system}

\begin{figure}
    \centering
		\includegraphics[width=.45\linewidth]{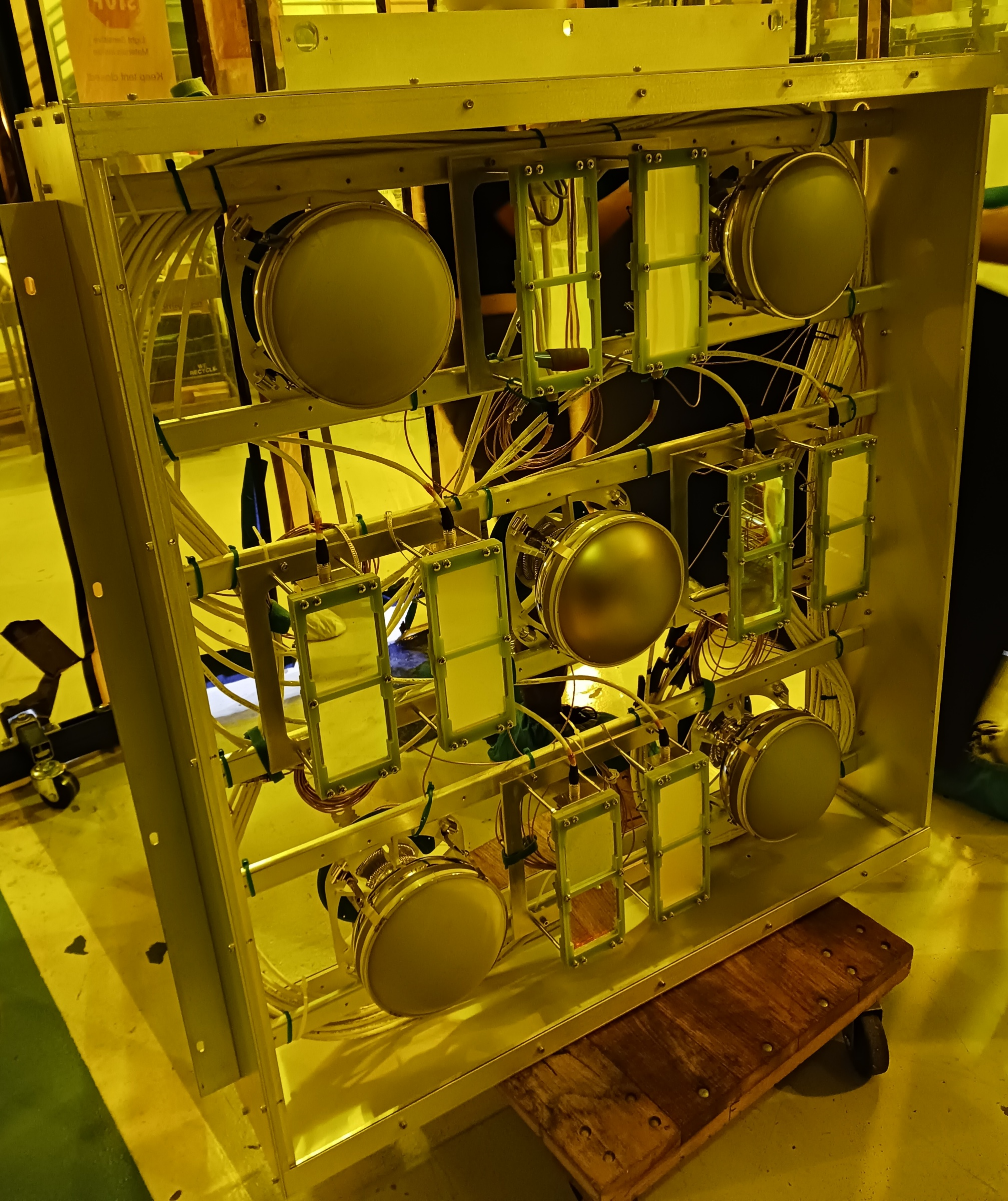}
	\caption{A SBND PDS box. The wavelength-shifter-coated PMTs are in the top and bottom rows; the single uncoated PMT is in the center. The coated rectangular-shaped X-ARAPUCAs are on the right of each pair, while the visible-sensitive ones are on the left.}
	\label{fig:pds_box}
\end{figure}

The SBND PDS uses two types of optical sensors, photomultiplier tubes and X-ARAPUCAs~\cite{Machado:2018rfb}, installed behind the anode wire planes and sensitive to visible or vacuum-ultraviolet (VUV) wavelengths. SBND uses 8-inch Hamamatsu R5912-MOD PMTs, 80\% of which are coated with a wavelength shifter that shifts the liquid argon VUV scintillation light at 128~nm into the visible range, peaking at 425~nm, where the PMTs have higher quantum efficiency. The remaining 20\% are left uncoated and therefore sensitive only to visible wavelengths. The PMTs serve as the primary components that trigger the LArTPC readout for neutrino and cosmic-ray interactions in SBND. In addition, the PDS is instrumented with 192 double-cell X-ARAPUCA modules, interleaved between the PMTs inside 24 \textit{PDS boxes}, one of which is shown in figure~\ref{fig:pds_box}. X-ARAPUCAs detect light via silicon photomultipliers (SiPMs) and also feature wavelength-shifting capabilities; half are coated to shift the liquid argon scintillation light, and the other half are left uncoated, similar to the PMTs. A summary of the number of sensors by type is given in table~\ref{tab:sbnd_light_sensors}. The combination of these two technologies gives SBND the largest surface LArTPC photo-sensitive coverage to date in a surface LArTPC, with nearly 15\% of the anode wall instrumented, corresponding to just under 4\% coverage of the entire inner detector surface. For comparison, the PDS of the MicroBooNE detector covers less than 4\% of its anode wall and less than 1\% of its whole inner detector surface~\cite{MicroBooNE:2016pwy}.

\begin{table}[ht]
\caption{Summary of the SBND PDS light sensitivities}
    \centering
    \begin{tabular}{|c|c|}
        \hline
        \textbf{Sensor} & \textbf{Light Sensitivity} \\
        \hline
        8-inch Hamamatsu PMTs & 96 sensitive to visible and VUV light \\
         & 24 sensitive to visible light only \\
        \hline
        X-ARAPUCA Modules (SiPMs) & 96 sensitive to visible and VUV light \\
         & 96 sensitive to visible light only \\
        \hline
    \end{tabular}
    \label{tab:sbnd_light_sensors}
\end{table}

\begin{figure}
	\centering
	\begin{tabular}{@{}c@{}}
		\includegraphics[width=.85\linewidth]{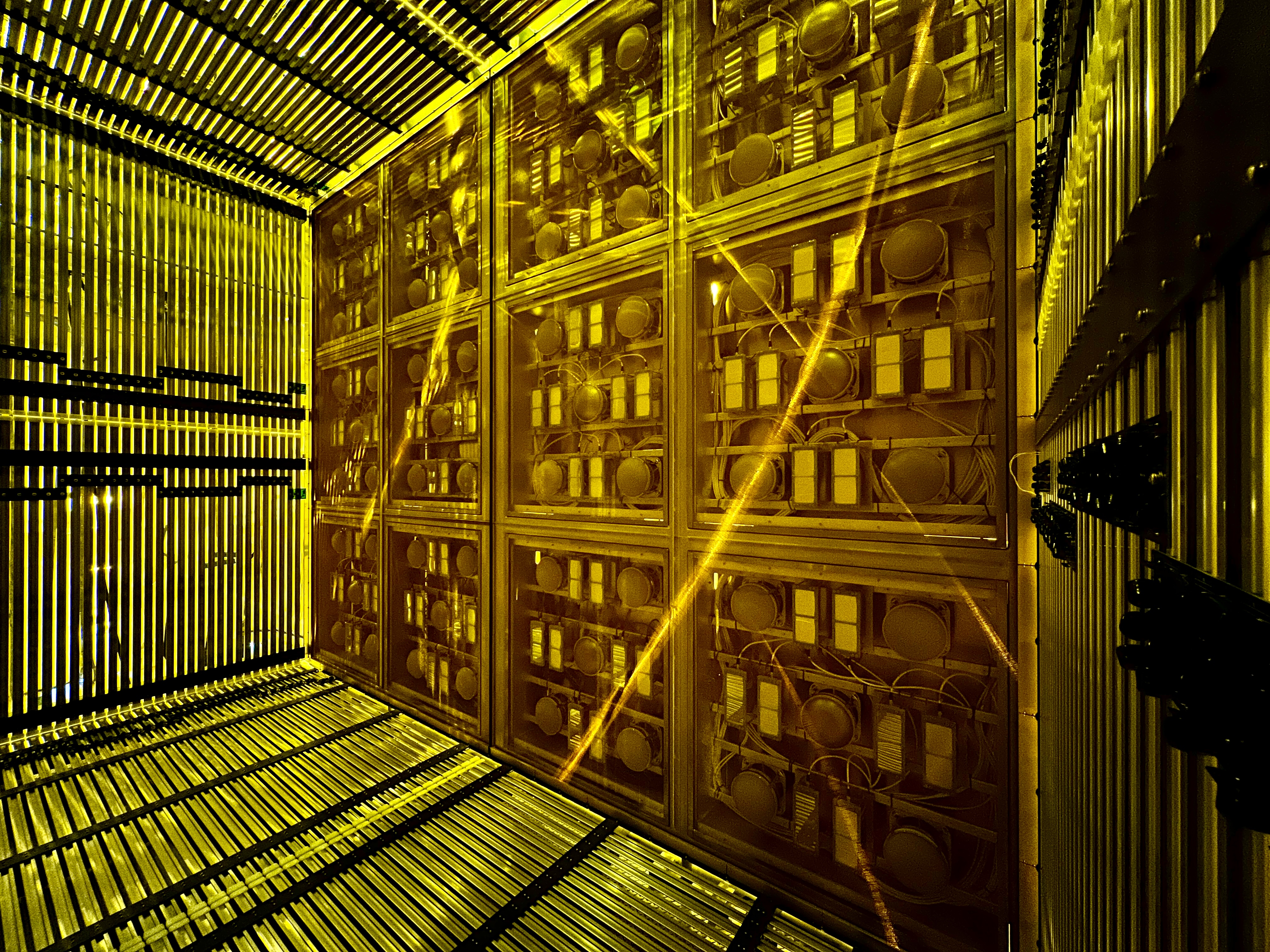} \\[\abovecaptionskip]
	\end{tabular}
	
	\vspace{\floatsep}
	
	\begin{tabular}{@{}c@{}}
		\includegraphics[width=.85\linewidth]{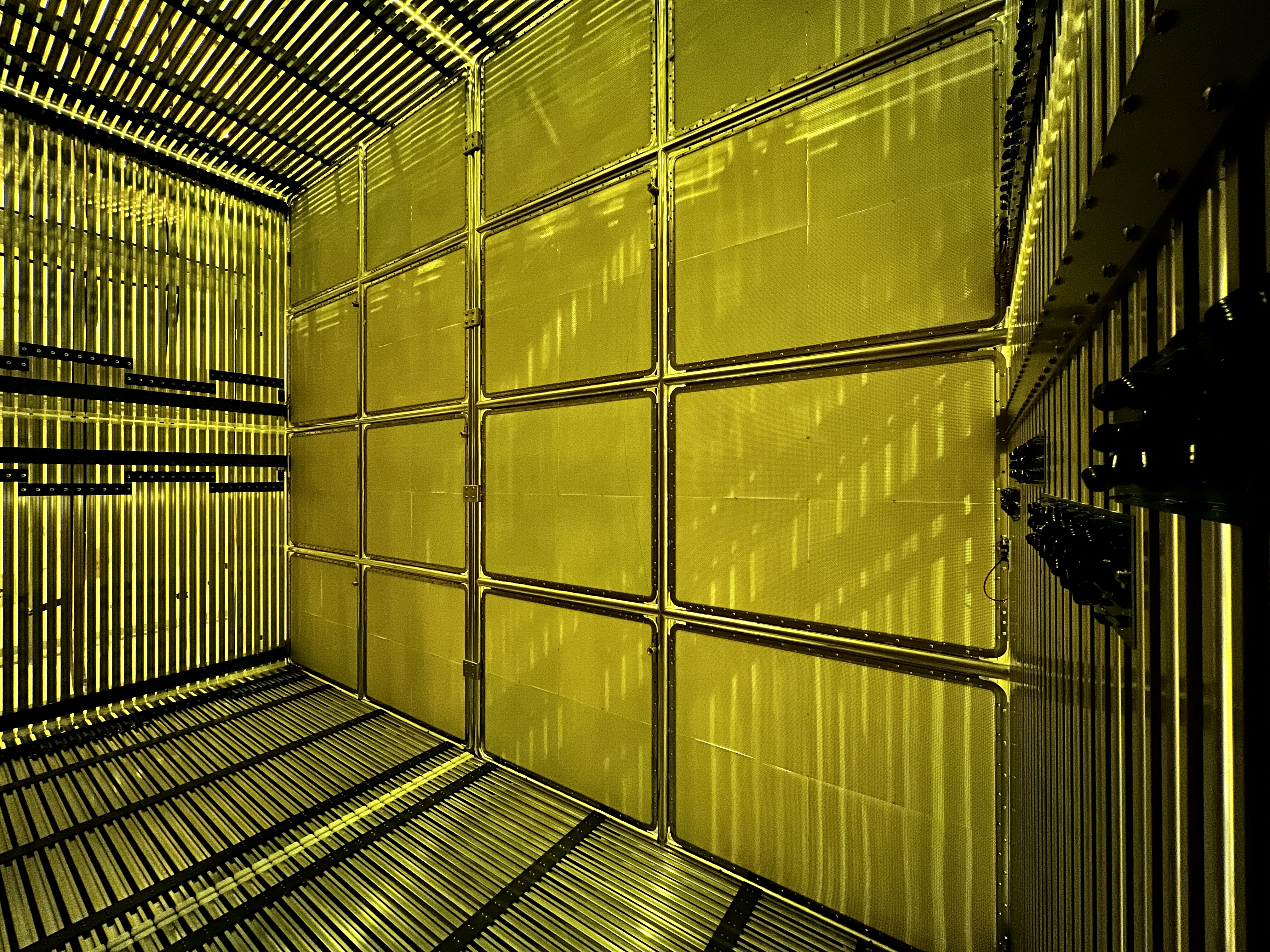} \\[\abovecaptionskip]
	\end{tabular}
	
	\caption{\textit{Top}: The fully installed SBND PDS wall located behind the anode plane wires. \textit{Bottom}: The cathode plane with the wavelength-shifting reflective plates placed behind a wire mesh.}\label{fig:sbnd_pds}
\end{figure}

As shown in the top panel of figure~\ref{fig:sbnd_pds}, each PDS wall is segmented into 12 individual PDS boxes. Each box contains five PMTs (four coated with wavelength shifter and one uncoated) and eight X-ARAPUCA modules (equally split between coated and uncoated units).

\begin{figure}
	\centering
		\includegraphics[width=.85\linewidth]{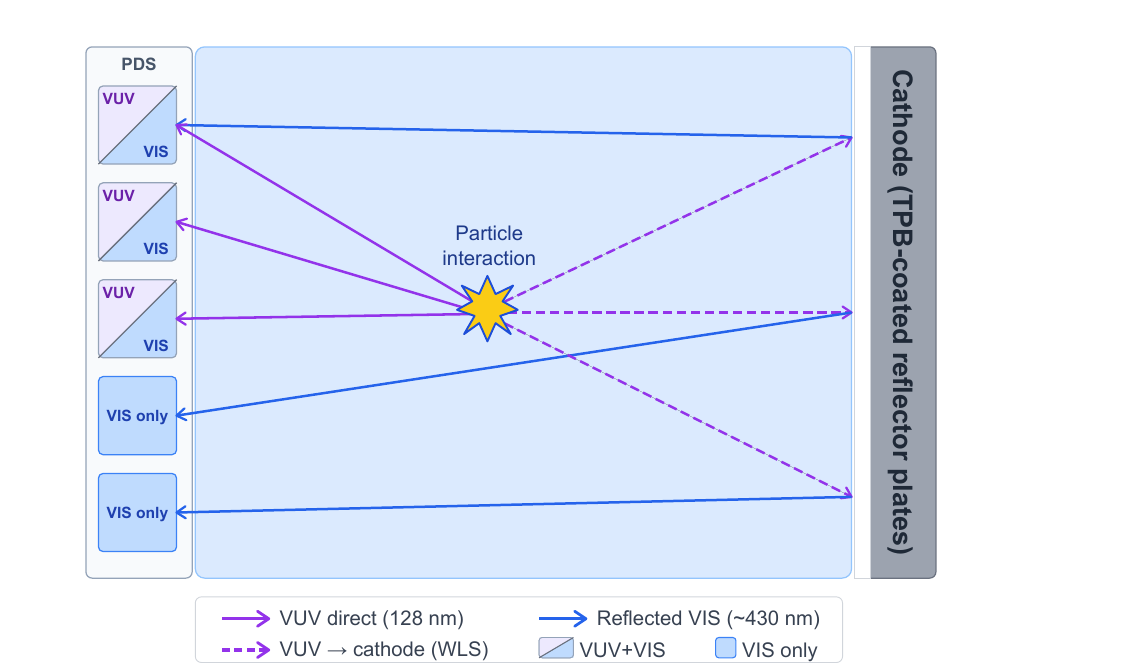}
	
	\caption{Cartoon representation of the VUV direct light and reflected visible (VIS) light from the wavelength-shifted cathode in SBND.}\label{fig:light_diagram}
\end{figure}

SBND incorporates the largest passive wavelength-shifting reflective detector wall deployed to date in a LArTPC, as shown in the bottom panel of figure~\ref{fig:sbnd_pds}. Positioned on the central double-sided cathode and forming the focus of this work, this wall enhances overall light-collection efficiency, since all PMTs and X-ARAPUCA modules—particularly the uncoated ones—are primarily sensitive to wavelength-shifted reflected visible light. A representative diagram of the two different types of light, direct VUV and reflected visible, and different detection capabilities of the PMTs and X-ARAPUCA are shown in figure~\ref{fig:light_diagram}. The addition of this reflective cathode increases the effective photon-sensitive area from 15\% to just over 30\% of the inner detector surface.

\section{Scintillation light \& motivation for enhancement}

Interaction in the liquid argon produces isotropic scintillation light, some of which reaches the photon detectors (PDs) directly. Conventional PDs, including PMTs and SiPMs, are effectively insensitive to liquid argon scintillation light emitted at 128~nm. This light lies in the vacuum ultraviolet (VUV) region of the electromagnetic spectrum, where most materials—including the glass used in PMTs and the materials used in detector walls—are opaque. The most common solution is to apply thin films of wavelength-shifting compounds to the outer surfaces of the PDs, so that incoming VUV photons are absorbed and re-emitted isotropically at longer wavelengths, typically in the blue region, where the PDs have high quantum efficiency. The organic compound tetra-phenyl butadiene (TPB) is the most widely used wavelength shifter in LArTPC experiments because of its high efficiency and reliability. It has been adopted in numerous detectors including ICARUS~\cite{SBND}, MicroBooNE~\cite{MicroBooNE:2016pwy}, WArP~\cite{BENETTI2008495}, DEAP~\cite{deap_detector}, LArIAT~\cite{Acciarri:2019wgd}, DarkSide~\cite{Zuzel_2017}, GERDA~\cite{GERDA:2022hxs}, LEGEND-200~\cite{Saleh:2026mon} and various DUNE-related prototypes~\cite{DUNE:2022ctp, Aimard:2020qqa}. SBND uses the TPB-coated reflective panels mounted on the cathode to recover both the VUV photons propagating away from the PDS and a small fraction of the visible light emitted backwards from the PMT coatings, thereby enhancing the total amount of light detected. In the SBND configuration, the photons reaching the cathode are wavelength-shifted and reflected toward the sensors. SBND’s versatile PDS is uniquely suited to observe both the direct VUV light and the secondary visible light reflected from the cathode. As listed in table~\ref{tab:sbnd_light_sensors}, not all sensors are sensitive to direct VUV light. The combined response enables differentiation between reflected and direct light and, therefore, precise position reconstruction in the drift direction, perpendicular to the PDS walls, using only light information~\cite{SBND:2024vgn}.

The most important impact of wavelength-shifting reflective plates is the significant increase in overall light yield and the improved uniformity of light collection across the SBND drift region~\cite{SBND:2024vgn}. This enhanced light yield has several positive implications for the detector’s performance and physics goals by allowing light-only calorimetry and position reconstruction, improving background rejection, and improving overall calorimetric information.

A prior small-scale proof-of-concept, employing wavelength-shifting reflective plates at the cathode, was successfully demonstrated in the LArIAT experiment~\cite{Acciarri:2019wgd} and motivated the production and installation of a similar system in SBND.

\section{Wavelength-shifting reflecting plate production}

Since the original SBND design did not include reflector plates, the implemented solution had to be adapted and integrated into the existing design. As mentioned, the cathode was initially designed with 16 transparent thick-mesh windows mounted on steel profiles. To accommodate the reflector plates within the existing design and constraints of the evaporation system used, 64 double-sided light-enhancing plates of 58.6~cm by 46.6~cm were designed—four per window—to fully cover the cathode surface, as shown at the bottom of figure~\ref{fig:sbnd_pds}. Each plate consists of a fiberglass impregnated with a 0.8~mm thick FR4 resin, laminated on both sides with 3M DF2000MA~\cite{3M}, a non-conductive polymeric film. DF2000MA exhibits >99\% specular reflection in the visible light range. The full production process of the wavelength-shifting reflective plates can be summarized in two major steps:
\vspace{-\topsep}
\begin{enumerate}[itemsep=-1mm]
\item Preparation of FR4 plates and lamination with reflective film;
\item Wavelength-shifter evaporation, inspection, and preparation for storage and shipping.
\end{enumerate}
\vspace{-\topsep}

The plate lamination process was performed at the Illinois Institute of Technology (IIT) in Chicago, USA. The evaporations were shared between the University of Manchester in Manchester, UK, and Universidade Estadual de Campinas (UNICAMP) in Campinas, Brazil.

\subsection{Reflector plate preparation and lamination}

\begin{figure}
	\centering
	\includegraphics[width=0.6\textwidth]{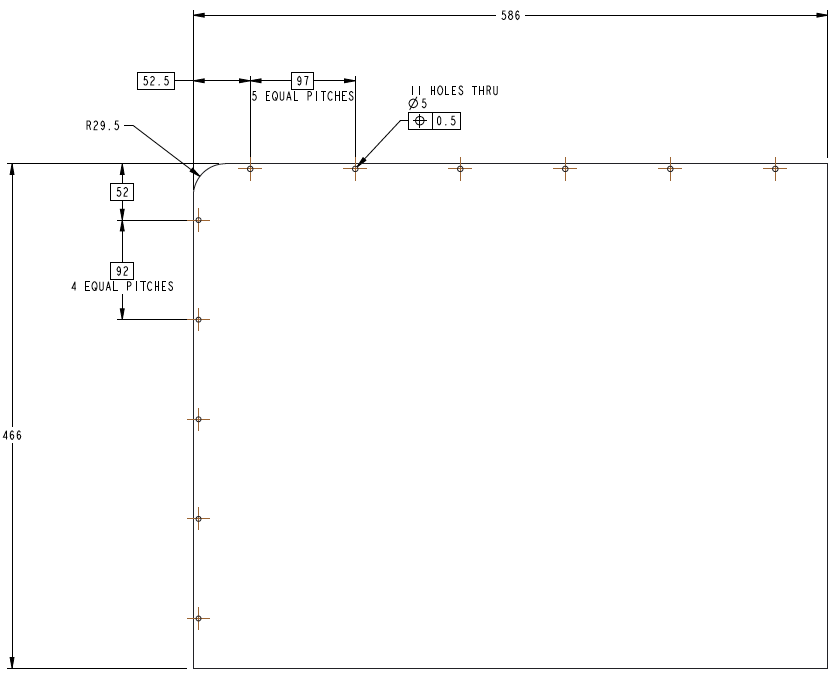}
	\caption{\label{fig:plates} Design sketch of a plate with indicated measurements in mm. Four of these plates are fitted into a single cathode window and mounted via screw holes.}
\end{figure}

\begin{figure}
		\centering
		\includegraphics[width=.49\linewidth]{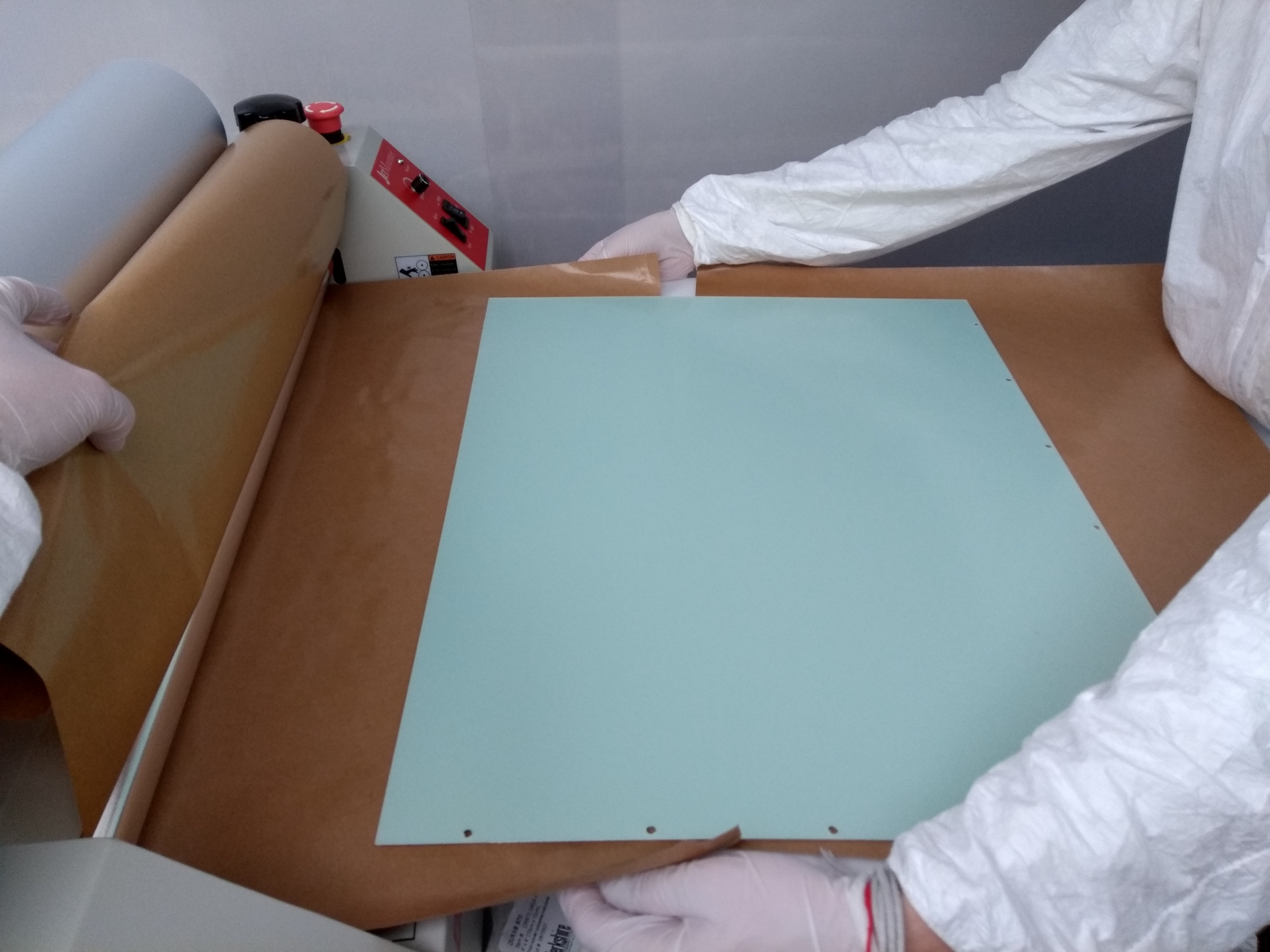}
		\includegraphics[width=.276\linewidth]{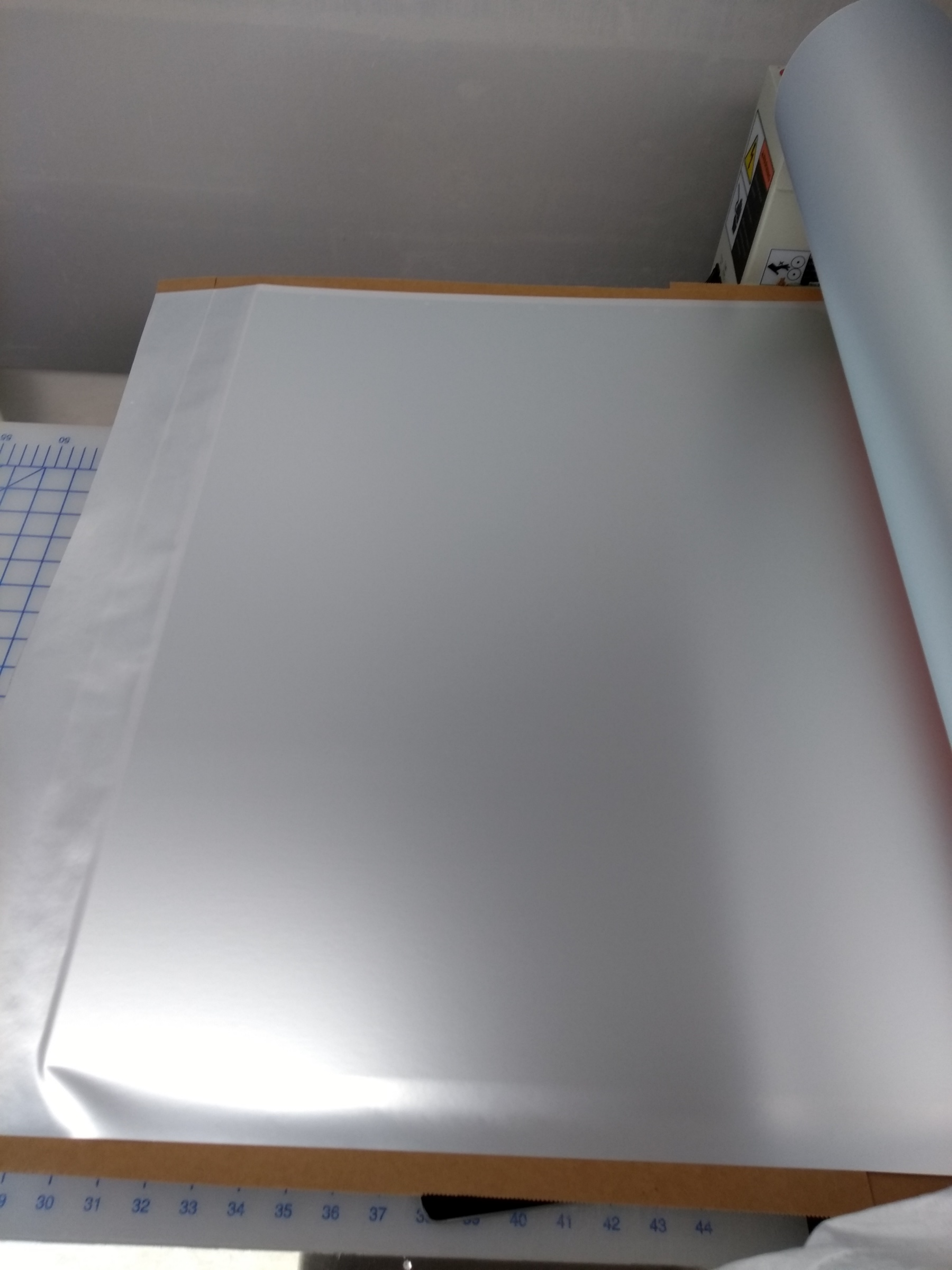}
	\caption{The plate lamination processes took place in IIT’s lamination cleanroom laboratory. \textit{Left:} Process of the plate entering the laminator. \textit{Right:} Plate coming out with the polymeric film now laminated on the surface. The excess reflector around the edges and holes had to be cut and removed by hand after the reflector was laminated on one side.}
	\label{fig:laminator_plates}
\end{figure}

The first two production steps were carried out in two phases: one during the summer of 2018 and the second in January 2019. Rectangular FR4 plates were procured and trimmed to the required dimensions, including drilling holes according to the design shown in figure~\ref{fig:plates}. The plate dimensions and hole locations were dictated by the layout of the cathode windows and the mounting structure. A Computer Numerical Control (CNC) machine template was used to ensure uniformity across all plates. Following cutting, each plate was hand-filed to remove rough edges and ensure that all holes were cleanly punched through. Once trimmed and cleaned, the plates were transferred to a cleanroom environment to mitigate debris contamination during lamination. The lamination was performed semi-automatically using a hot laminator. Each plate was laminated one side at a time using the reflective film. This film had an adhesive backing that adhered to the FR4 surface during lamination, while a matte protective layer on the opposite side remained in place until the evaporation stage. Figure~\ref{fig:laminator_plates} shows an example of a plate entering the laminator (left) and exiting (right) with the reflective film applied. Before laminating the second side, the excess film was trimmed and the holes manually cleared with a scalpel, as shown in figure~\ref{fig:laminator_plates} (bottom). The plate was then reinserted into the laminator to repeat the process on the second side. After lamination, each plate underwent a visual inspection to check for deformation or poor adhesion. The approved plates were then carefully packed and shipped to the evaporation site. More details of the lamination operations performed with this technique, which was also used to fabricate reflector elements for the PROSPECT~\cite{PROSPECT:2018dnc} and CCM-200~\cite{CCM:2025kal} experiments, can be found in Ref.~\cite{PROSPECT:2019enz}.

\subsection{Wavelength shifter evaporation, inspection, and storage}

\begin{figure}
	\centering
    \includegraphics[width=.49\linewidth]{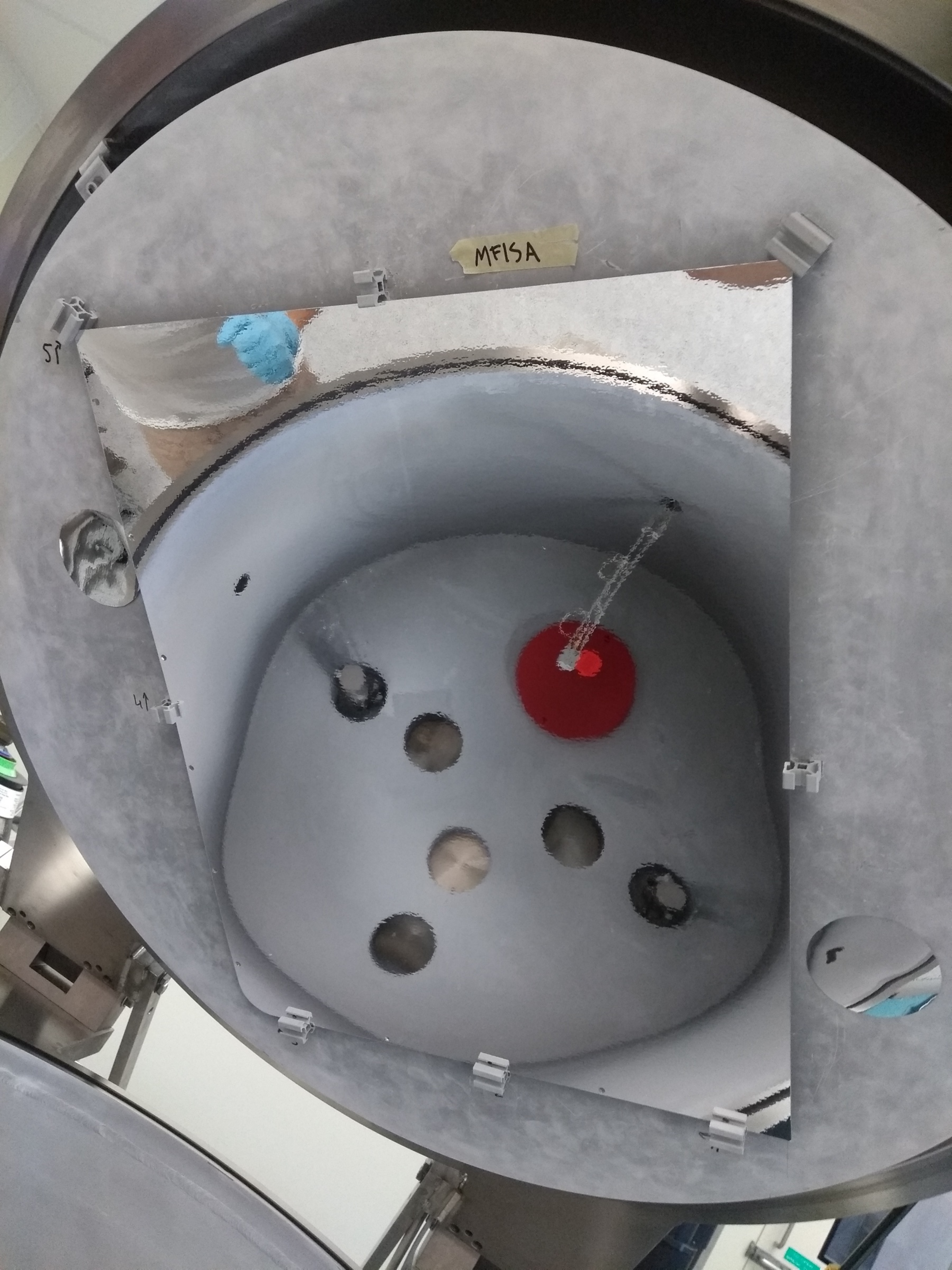}
    \includegraphics[width=0.49\textwidth]{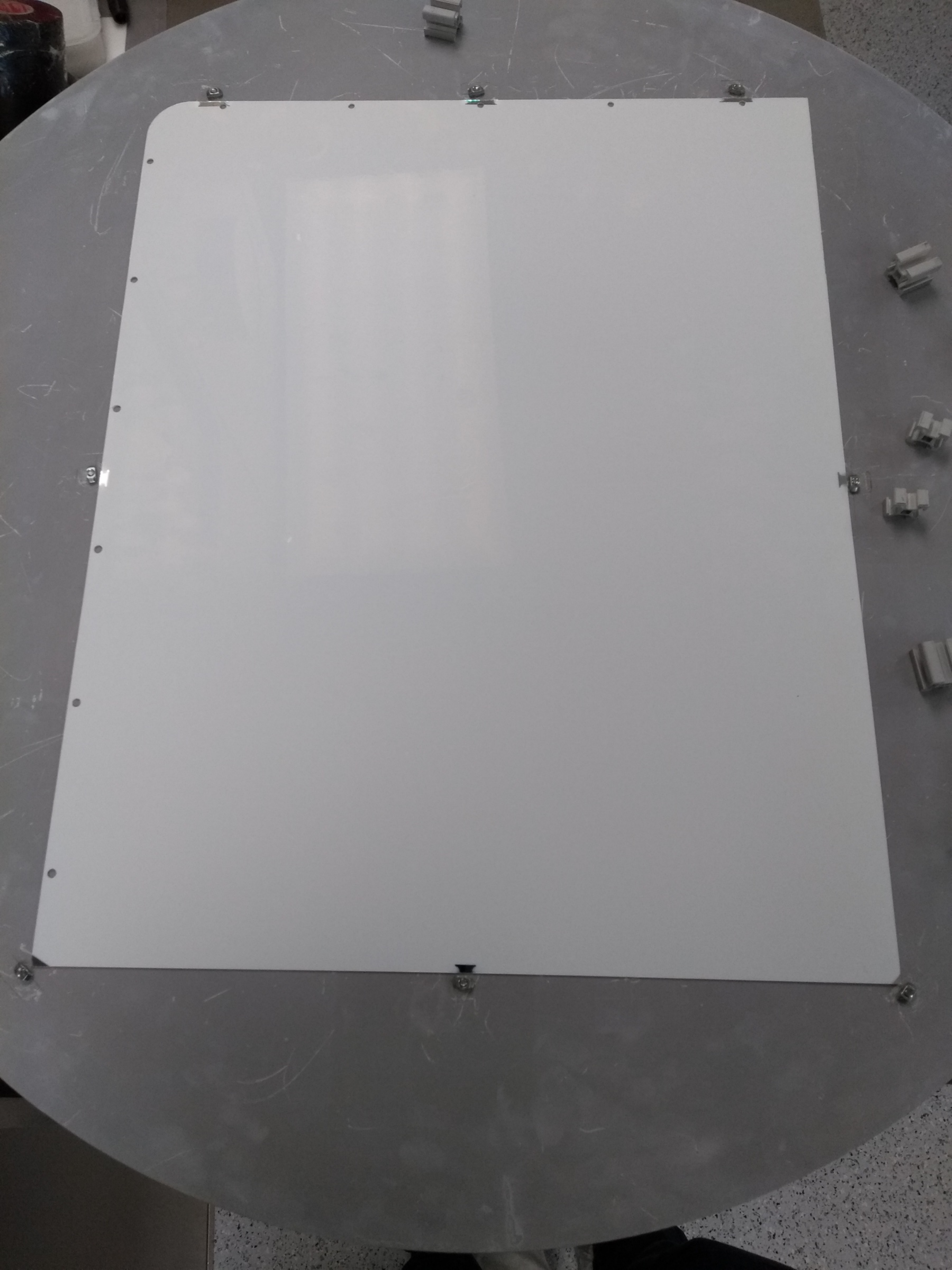}
	\caption{\label{fig:evaporated_foils}\textit{Left:} The disc with the mounted plate placed in the evaporator ready to be evaporated on. \textit{Right:} Example of one of the finished plates for SBND with a TPB layer deposited on top of the reflective film.}
\end{figure}

As previously mentioned, TPB was chosen as the wavelength shifter for the light enhancers due to its high performance and reliability in converting VUV photons to wavelengths matching the peak quantum efficiency of the cryogenic PMTs used in SBND. Thin layers of approximately 300~$\upmu$g/cm\textsuperscript{2} were evaporated onto the reflective films shown in figure~\ref{fig:evaporated_foils} (left), resulting in a characteristic pearl-white appearance, as shown in figure~\ref{fig:evaporated_foils} (right).

The TPB coatings were applied using a low-temperature physical vapour deposition (PVD) technique. In this method, TPB powder is placed inside small crucibles and heated above its sublimation point, allowing it to vaporize and deposit uniformly onto the substrate positioned directly above.

\begin{figure}
	\centering
	\includegraphics[width=0.6\textwidth]{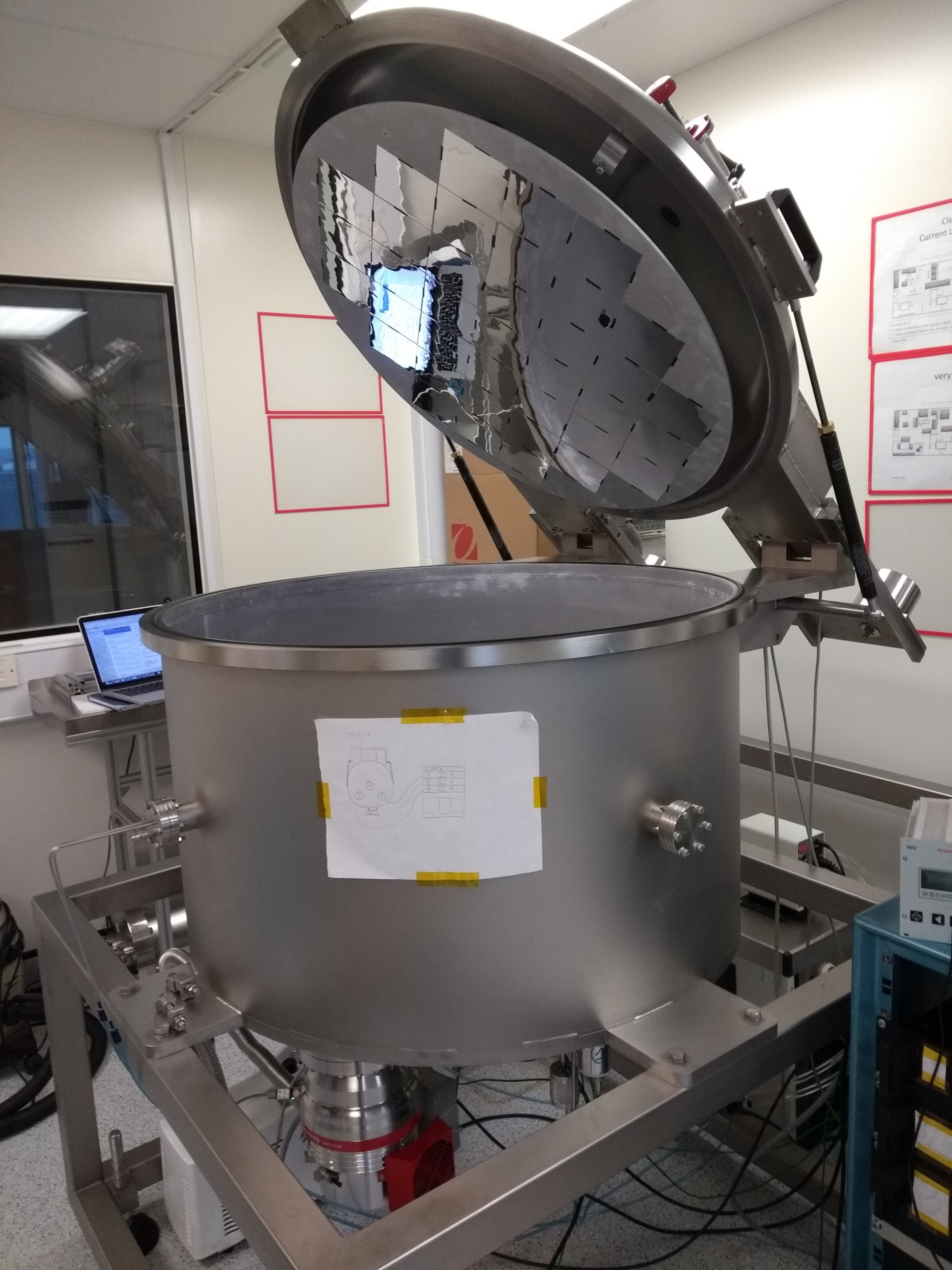}
	\caption{\label{fig:evaporator}The evaporation chamber used for the evaporation of TPB. In the photo, calibration small samples from section~\ref{sec:calib} were attached to the rotating disc mounted under the lid, rather than to the SBND plate, as shown in figure~\ref{fig:evaporated_foils}.}
\end{figure}

The entire evaporation process takes place in a high-vacuum environment to minimize contamination from impurities, which could compromise both the performance and adhesion of the TPB layer. A vacuum chamber—referred to as the evaporator—was used for this purpose, as shown in figure~\ref{fig:evaporator}. The substrates were mounted on a rotating disc affixed to the chamber lid. This rotation ensures uniform deposition over the large surface area of the reflector plates.

\begin{figure}
	\centering
	\includegraphics[width=0.4\textwidth]{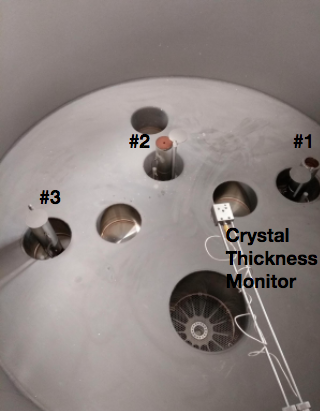}
    \includegraphics[width=.4\textwidth]{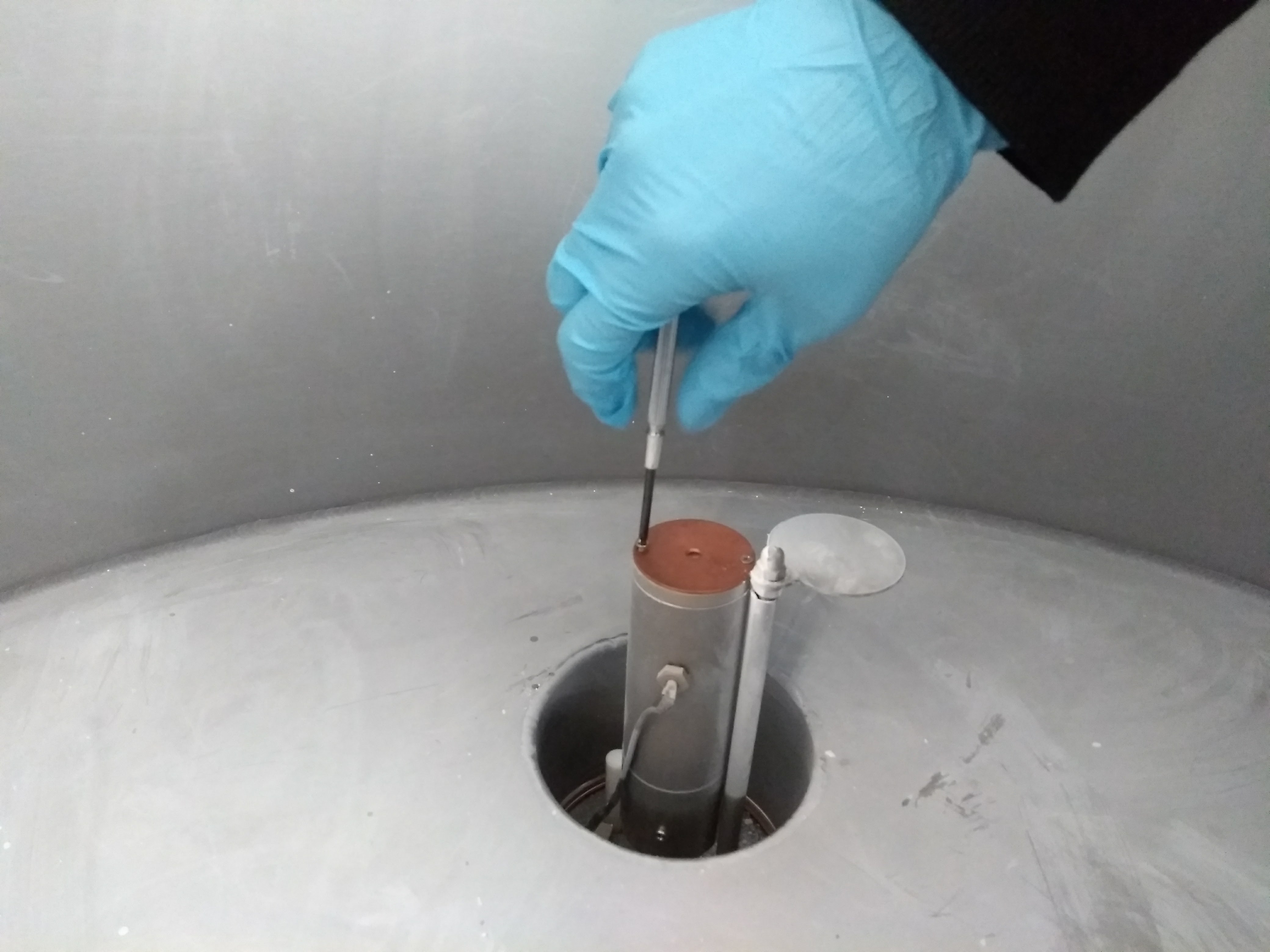}
	\caption{\label{fig:crucible}\textit{Left:} The inside of the evaporation chamber with the locations of the three different crucibles (labelled \#1, \#2 and \#3 respectively), the turbo pump feed-through, and the crystal thickness monitor. \textit{Right:} The crucible with a pinhole copper lid being screwed in to reduce and stabilize the evaporation rate.}
\end{figure}

At the base of the chamber, three crucibles were installed, along with a crystal thickness monitor, as shown in figure~\ref{fig:crucible}. The side crucibles, \#1 and \#3, were primarily used due to their superior coating performance on large substrates, resulting from their off-center placement, which allows for more uniform coverage as the disc rotates. Each crucible contained a small aluminum cup filled with 5.0~g of TPB. These cups were covered with pinhole copper masks to regulate the evaporation rate and prevent the TPB from boiling over, as shown in figure~\ref{fig:crucible}. Heating was provided by internal coils inside the crucibles, with power and temperature controlled via external sensors and a regulated power supply.

Since each plate required coating on both sides, it could not be mounted directly onto the rotating disc, as this would risk damaging the previously applied reflector and TPB layers. To minimize contact with the disc and avoid masking large areas of the reflector, eight short Bosch-profile support structures were strategically positioned to hold each plate securely in place without sagging or falling. Figure~\ref{fig:evaporated_foils} (left) shows a photograph of a plate supported by these profiles while mounted on the evaporator. A set of T-nuts was employed to lock the profiles in place. This setup allowed quick loosening of selected supports, allowing the reflector plates to be easily inserted or removed from the evaporation setup without disturbing the rest of the structure. The holding system covered less than 0.2\% of the total reflector surface area and ensured a stable, level platform for uniform coating throughout the plate, as shown in figure~\ref{fig:evaporated_foils} (right).
 
\subsubsection{Evaporation process}

\begin{figure}
	\centering
	\includegraphics[width=0.85\textwidth]{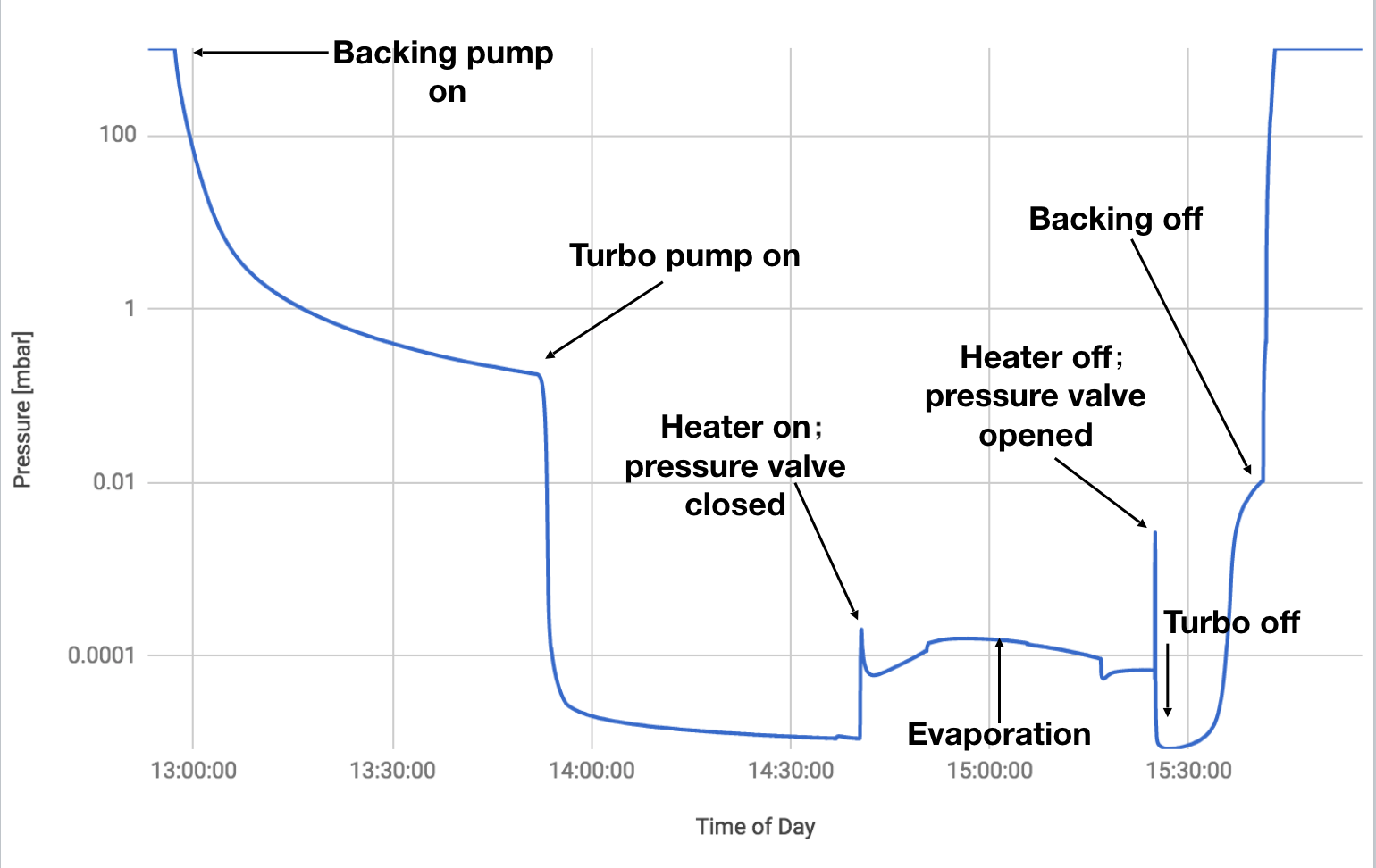}
	\caption{\label{fig:pressurecycle}Typical pressure cycle during an evaporation from closing the lid to opening it back up again. A low vacuum of 10$^{-5}$~bar was typically targeted to minimize the risk of contaminants during evaporation.}
\end{figure}

The complete evaporation process lasted approximately 2.5 hours, which included evacuating the chamber using Pfeiffer ACP 28 multi-stage roots vacuum~\cite{roughpump} (roughing) and Pfeiffer HiPace 700~M turbo-molecular pumps~\cite{turbopump}, performing the evaporation, and venting the chamber to allow for substrate exchange. Initially, a mid-to-low vacuum of approximately $10^{-2}$~mbar was reached using a roughing pump (serving as the backing pump). The process was followed by a high-vacuum stage, using a turbo-molecular pump to achieve pressures of at least $10^{-5}$~mbar, at which point evaporation could begin. Each evaporation followed a consistent cycle, summarized by the pressure profile shown in figure~\ref{fig:pressurecycle}.

\begin{figure}
	\centering
	\includegraphics[width=0.8\textwidth]{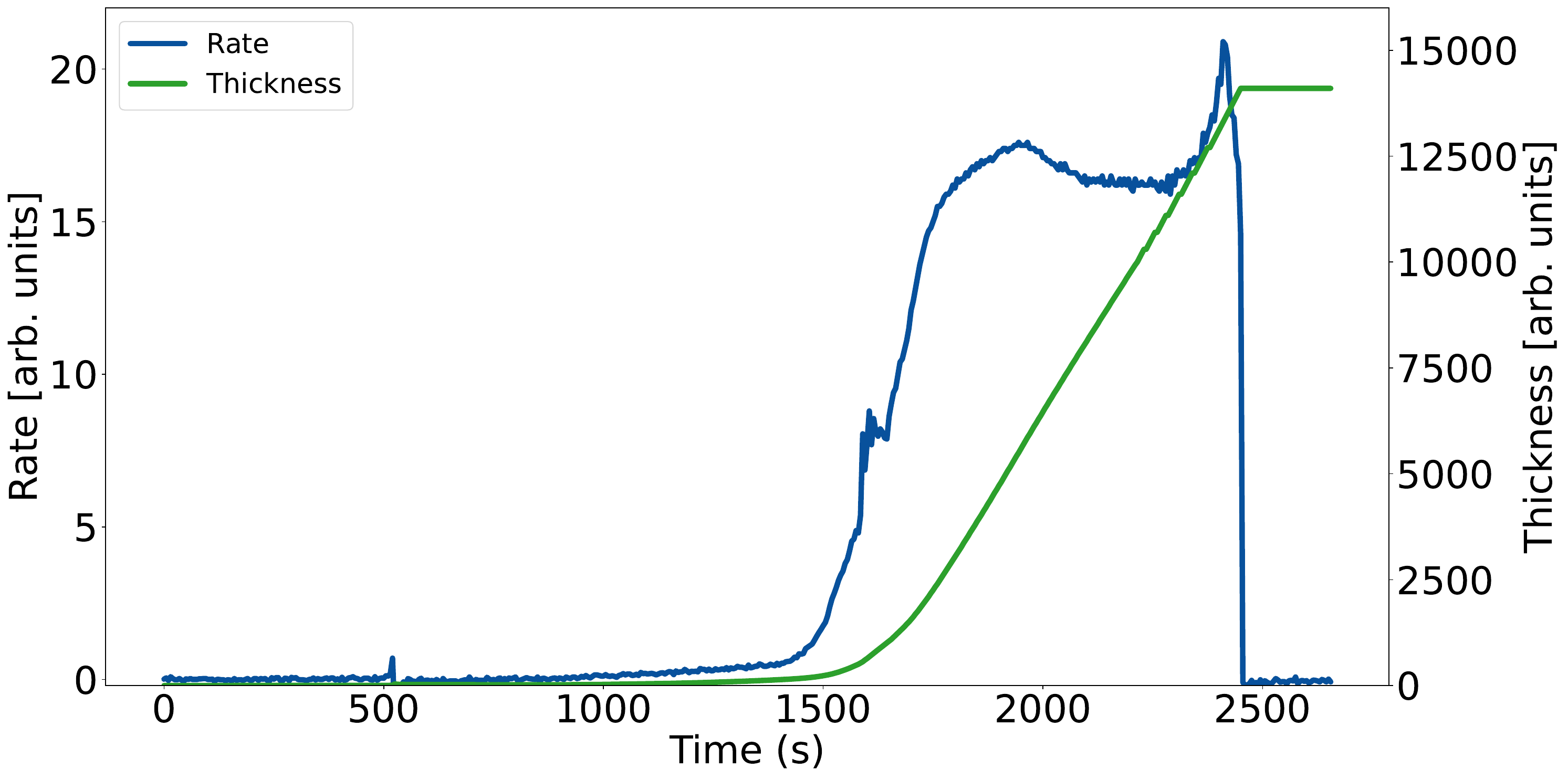}
	\caption{\label{fig:evaporationcycle}Relative rate and thickness over time for a typical evaporation cycle. The blue curve represents the evaporation rate, which slowly rises as the crucible temperature approaches the set temperature that maximizes evaporation; the green curve is the integrated film thickness.}
\end{figure}

After the desired vacuum pressure was achieved, typically about an hour after the turbo pump was activated, the plate rotation was started, and the crucible’s heating coil was powered with the temperature set to 245~$^\circ$C. This temperature was chosen to maintain the TPB evaporation within its optimal range of 220–250~$^\circ$C. The set-point of 245~$^\circ$C was determined through multiple test runs under similar conditions to produce high-quality evaporations, as indicated by visual inspections. The evaporation rate was carefully monitored using a quartz crystal thickness monitor, as shown in figure~\ref{fig:crucible}. This device measures the deposition rate, typically in angstroms per second, by detecting changes in the crystal's oscillation frequency caused by TPB molecules depositing on its surface. The crystal oscillates at MHz frequencies, and the frequency shift can be translated into rate and thickness using the physical properties (e.g., density and compressibility) of the evaporated material. Since dedicated material constants for TPB were unavailable, the monitor readings served primarily as relative references for comparing different evaporations and determining their start and end times. The evaporator was calibrated using a series of dedicated measurements, as detailed in Section~\ref{sec:calib}, to ensure the desired coating thickness was achieved. Figure~\ref{fig:evaporationcycle} shows a typical evaporation cycle, with the deposition rate plotted in blue along with the relative thickness over time. As the crucible temperature rose and the TPB began to boil, the evaporation rate gradually increased until it stabilized when a uniform temperature was reached. Once all the TPB in the cup had evaporated, the rate dropped sharply back to zero.

\begin{figure}
  \centering
   \includegraphics[width=0.2466\textwidth, trim={18cm 0 35cm 0},clip]{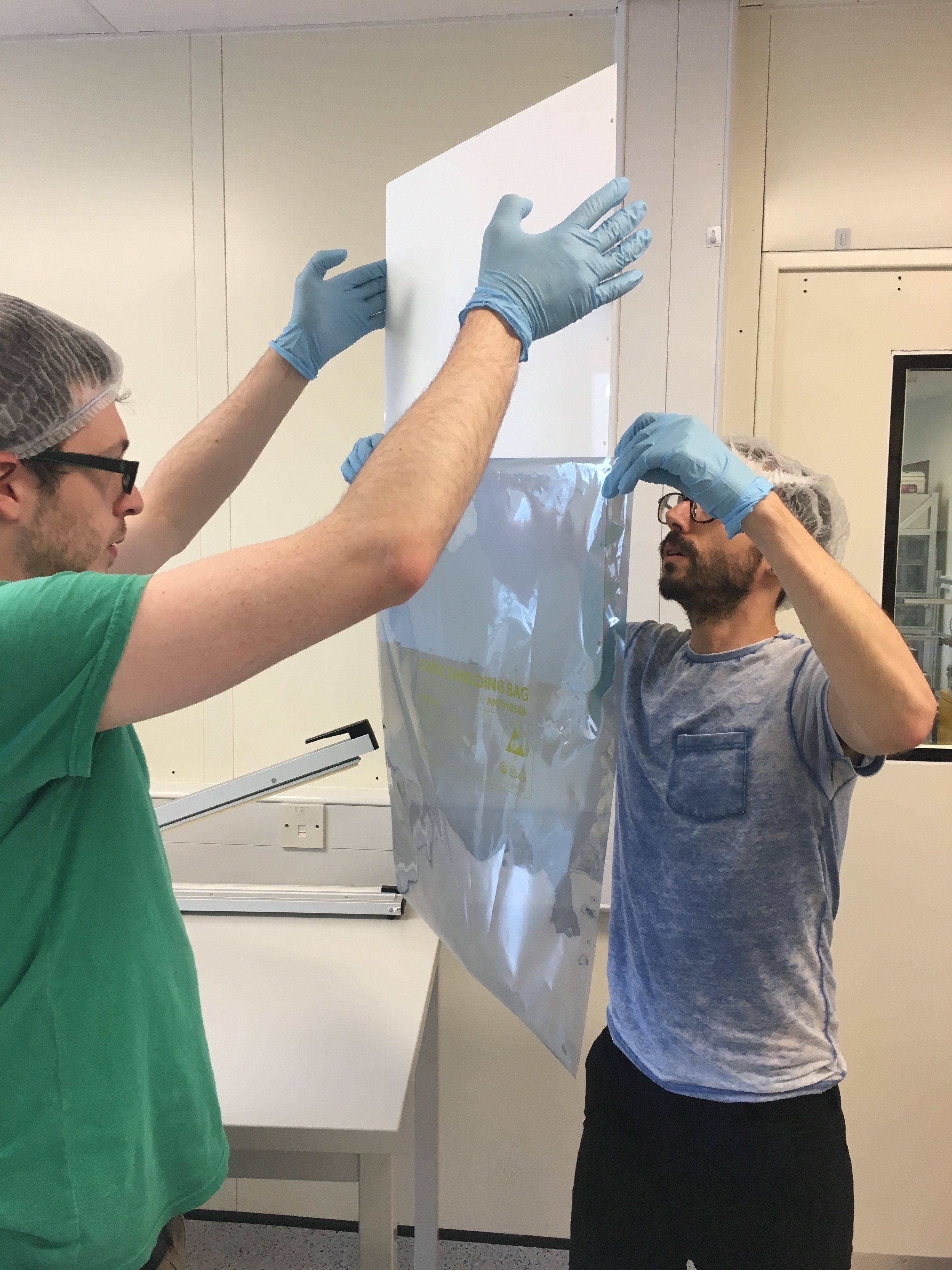}
   \includegraphics[width=0.49\textwidth]{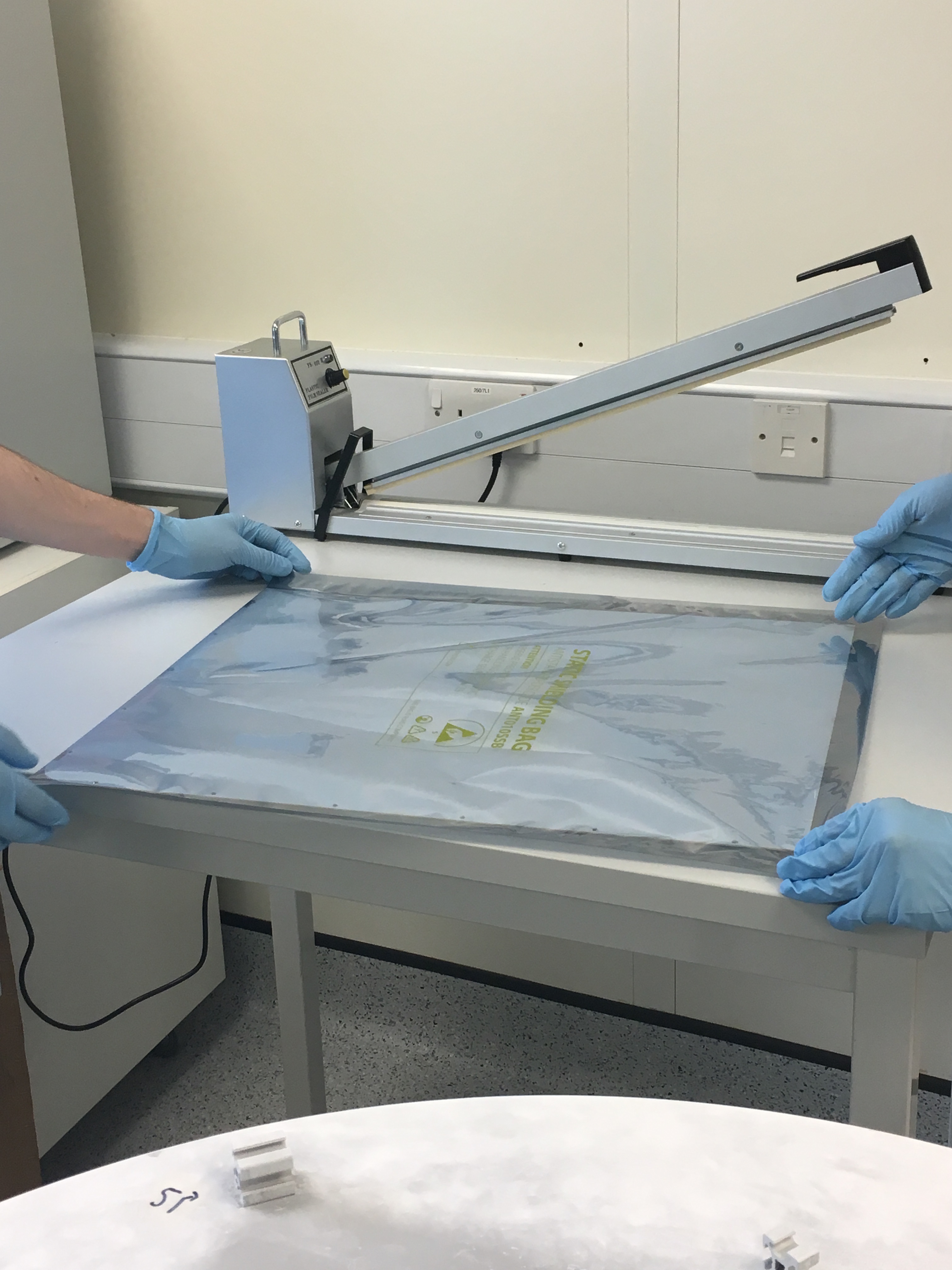}
 \caption{\textit{Left}: The insertion process of a completed plate into a protective bag where one individual keeps the bag open, while the second slowly inserts the plate into the bag, holding it on its side and avoiding contact with the bag’s surfaces. \textit{Right}: The bags are sealed using the heat sealer. One side of the bag is sealed to prevent the plate from moving laterally, while the bag opening is partially sealed before the bag is filled with nitrogen gas. Once it is filled, the opening is fully sealed to prevent the nitrogen from escaping.}
\label{fig:sbnd_bag_sealer}
\end{figure}

Once the deposition rate stabilized at zero for several minutes, confirming that no TPB remained in the crucible cup, both the crucible heater and the turbo pump were turned off to prepare for venting the chamber. While the turbo pump slowed down safely, preparations for the next evaporation cycle began, such as filling a new cup with TPB powder. After the turbo pump had fully stopped, the backing pump was powered off, and the vacuum was softly broken by slowly injecting pure nitrogen gas into the chamber. This process minimized exposure to humidity and oxygen, protecting the freshly deposited film as it cooled down. Once atmospheric pressure was reached, the chamber lid was opened, and the rotating disc was removed and placed on a workbench. The uniformity of the evaporation was visually verified to ensure that the wavelength shifter covered the plate. The plate was either flipped to coat the other side or swapped with a new one. The plate was carefully removed by loosening a few Bosch-profile supports just enough to slide it off. During this process, only the edges of the plate were handled to avoid damaging either coated surface. If both sides of the plate had been coated, it was immediately placed in a large UV-protective, anti-static bag, which was then filled with nitrogen and sealed with a heat press. This procedure was performed as quickly as possible to minimize exposure to ambient air and light. Figure~\ref{fig:sbnd_bag_sealer} shows a plate being inserted into the bag and partially sealed before filling. In addition to limiting humidity and airborne contaminants, nitrogen-filled bags provided cushioning to protect the plates during storage and transport. The sealed bags were then stored in a dark box, ready for shipment. This entire process was repeated for each plate until the required number, including spares, was produced.

In summary, 81 plates were produced, 68 of them qualified as good to be installed and 13 as spare, with a total required of 64, as summarized in table~\ref{tab:summary_evap}.

\begin{table}[h]
	\caption{Summary of evaporation campaigns with plates divided into two categories for installation.}
	\centering
	\def\sym#1{\ifmmode^{#1}\else\(^{#1}\)\fi}
	\begin{tabular}{l*{4}{c}}
		& \multicolumn{1}{c}{Manchester}  & \multicolumn{1}{c}{Unicamp} & \multicolumn{1}{c}{Manchester}  & \multicolumn{1}{c}{\textbf{Total}} \\
		& \multicolumn{1}{c}{(Aug. 2018)}  & \multicolumn{1}{c}{(Dec. 2018)} & \multicolumn{1}{c}{(Feb. 2019)}  & \multicolumn{1}{c}{} \\
		\midrule
		Primary    &  22    &  28 & 18  & \textbf{68}    \\
		Spare    &  9    &  3 & 1   & \textbf{13}    \\
		\bottomrule
		Total    &  31  &  31  & 19   & \textbf{81}   \\
	\end{tabular}

	\label{tab:summary_evap}
\end{table}

\subsubsection{Calibration of the evaporator}{\label{sec:calib}}

\begin{figure}
	\centering
	\includegraphics[width=0.49\textwidth]{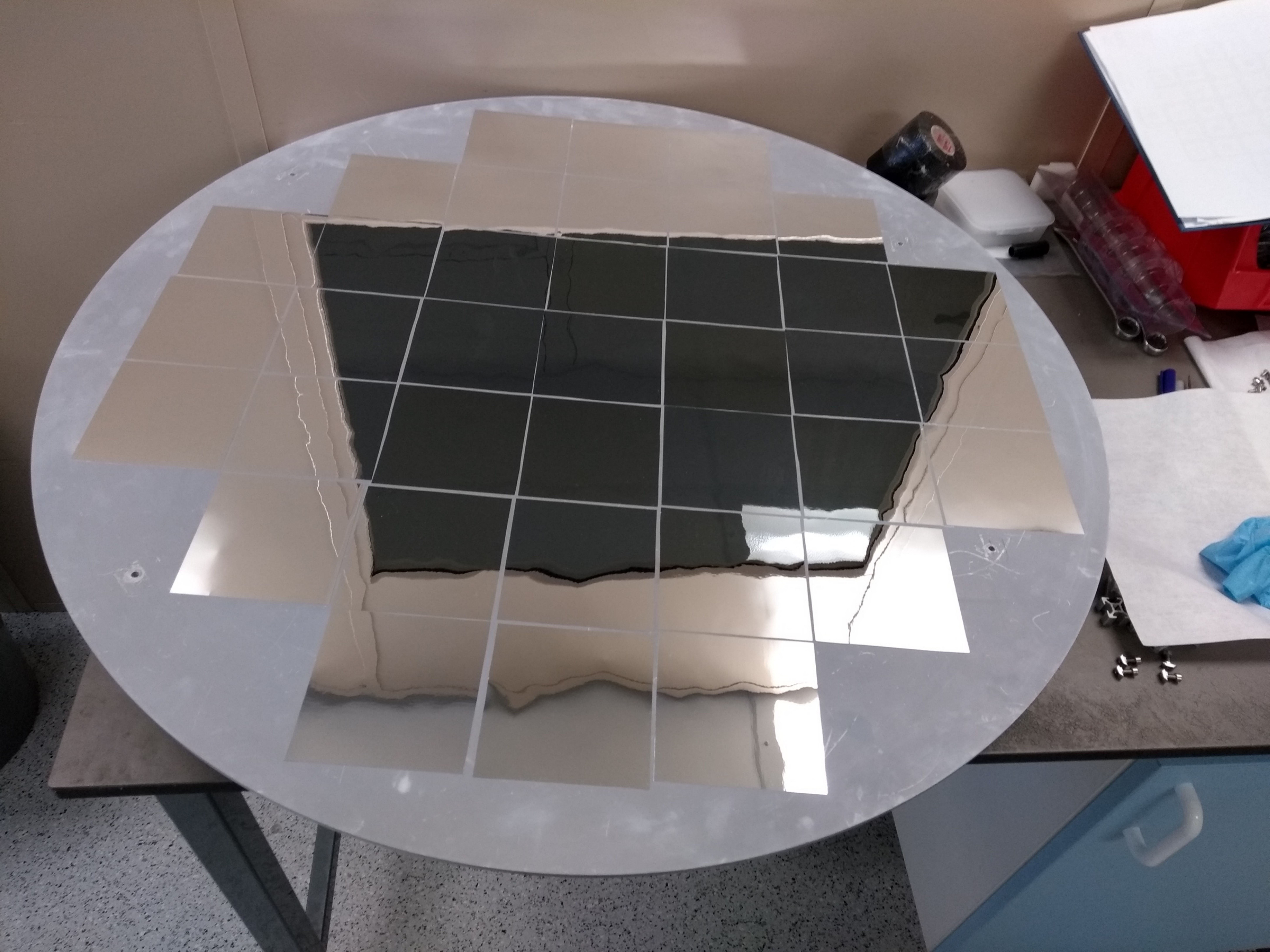}
	\includegraphics[width=0.49\textwidth]{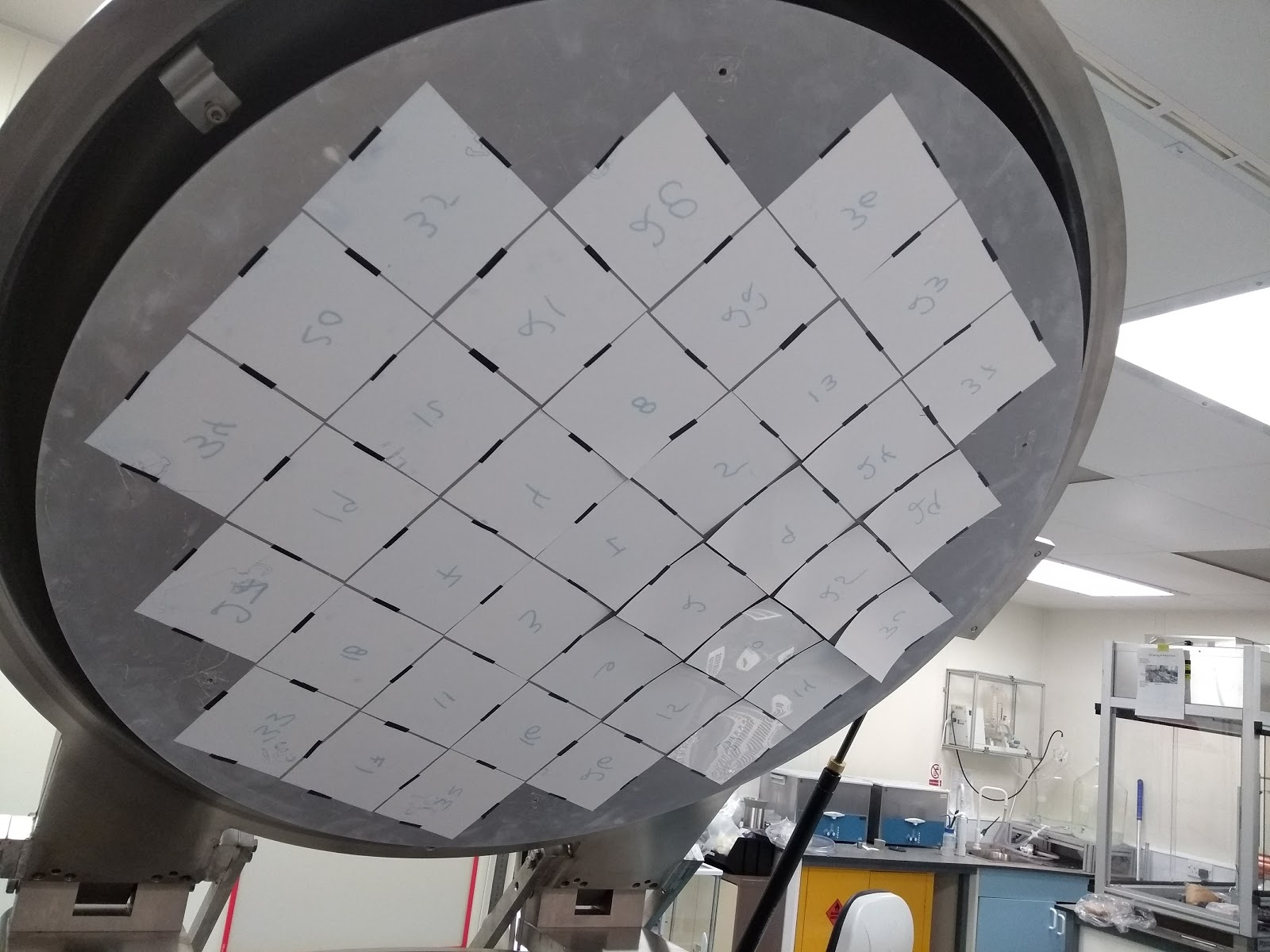}
	\caption{\label{fig:calib}$\it{Left}$: The calibration samples before the evaporation, placed at different radii. $\it{Right}$: Calibration samples after the evaporation. Each sample had dimensions of 10.5~cm x 10.5~cm.}
\end{figure}

\begin{figure}
	\centering
	\includegraphics[width=0.99\textwidth]{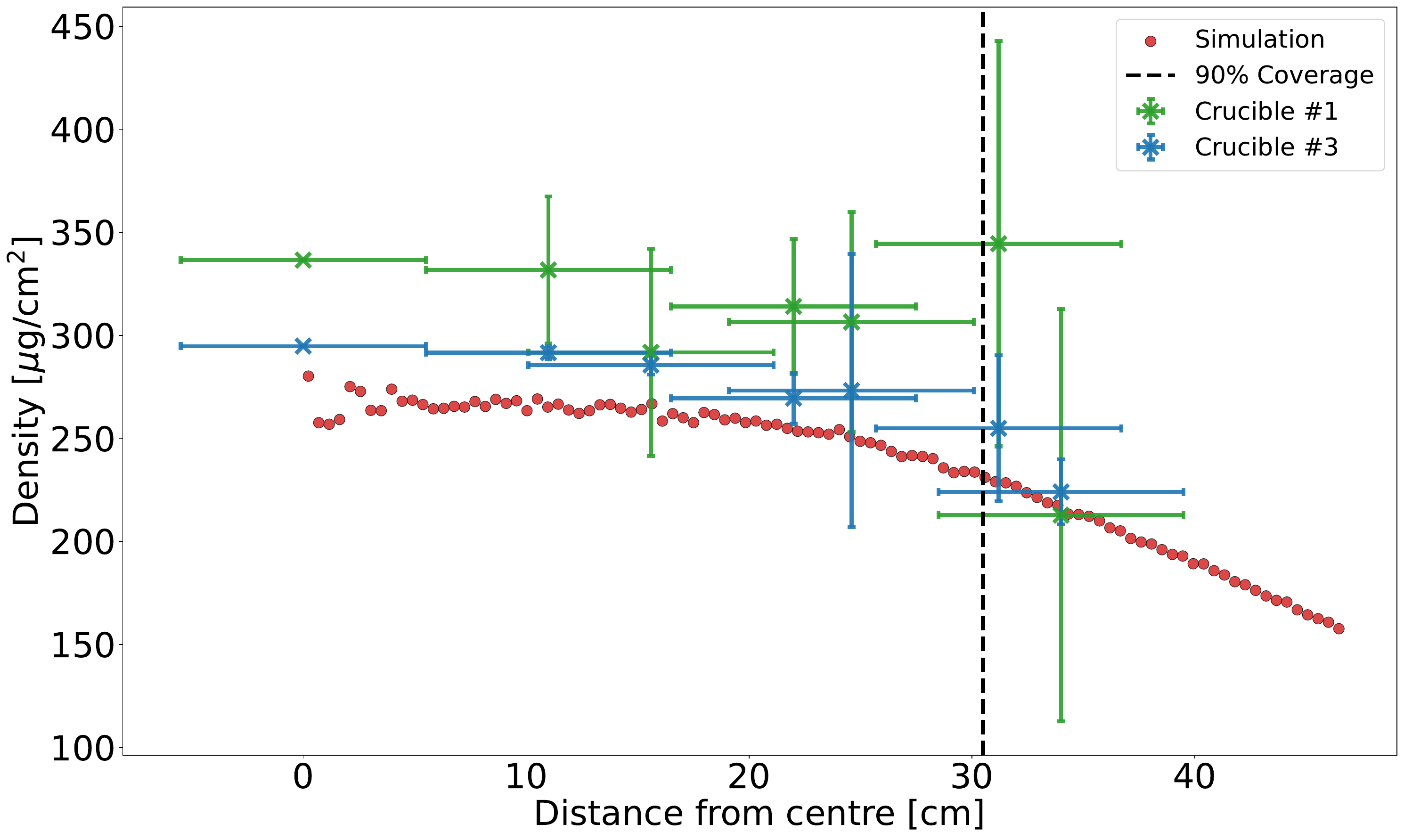}
	\caption{\label{fig:evaporationcalib}Calibration results of crucible \#1 and \#3 with 37 samples at different distances from the center. The distance is determined from the difference between the center of the disc and the center of the square. The horizontal bars account for the distance between the other extremities of the square to the central point. In contrast, the vertical ones are based on the standard deviation for different samples at the same distance from the center. The results show that the samples have an average mass density of approximately 325~$\upmu$g/cm$^{2}$ over 90\% of the plate area.}
\end{figure}

Achieving uniform coatings on the large reflector plates is challenging. Therefore, dedicated studies were conducted to determine the surface density distribution across the full rotating disc. A nominal coating thickness of 300~$\upmu$g/cm$^{2}$ was selected based on previous studies~\cite{Francini:2013lua}, which showed that thicknesses between 200~$\upmu$g/cm$^{2}$ and 1000~$\upmu$g/cm$^{2}$ yield maximal VUV light conversion efficiency with minimal light absorption by the TPB layer. This thickness also minimizes TPB waste due to typical evaporation inefficiencies. A geometric simulation was performed to predict the coating surface density as a function of the initial TPB quantity in the crucible and the distance between the crucible and the substrate. The results were compared with calibration runs, which involved placing 37 small reflector samples, each measuring 10.5~cm per side, at different locations across the disc to maximize surface coverage. Figure~\ref{fig:calib} shows an example calibration run before and after evaporation, where the samples covered an area larger than the reflector plates. The samples were weighed before evaporation. After coating, each sample was carefully peeled off and weighed on a sensitive scale housed in a draft-free enclosure to minimize airflow effects that could bias measurements. The surface density of the TPB deposition was calculated by dividing the measured mass difference by the sample's surface area. Figure~\ref{fig:evaporationcalib} presents the measurements from the two calibration runs using two crucibles that were symmetrically across from the center of the disc, along with simulation results assuming 5.0~g of TPB powder in the crucible. The vertical uncertainty bars correspond to the sample standard deviation of measurements at fixed radial distances from the center of the disc. Due to the limited number of samples available at each radius, the quoted uncertainties are sensitive to statistical fluctuations. These calibration results confirmed that the coating thickness across the plates was more or less uniform for at least 90\% of the area and, more importantly, within the ideal density range when using the proposed quantity of TPB.

\begin{figure}
	\centering
	\includegraphics[width=0.65\textwidth]{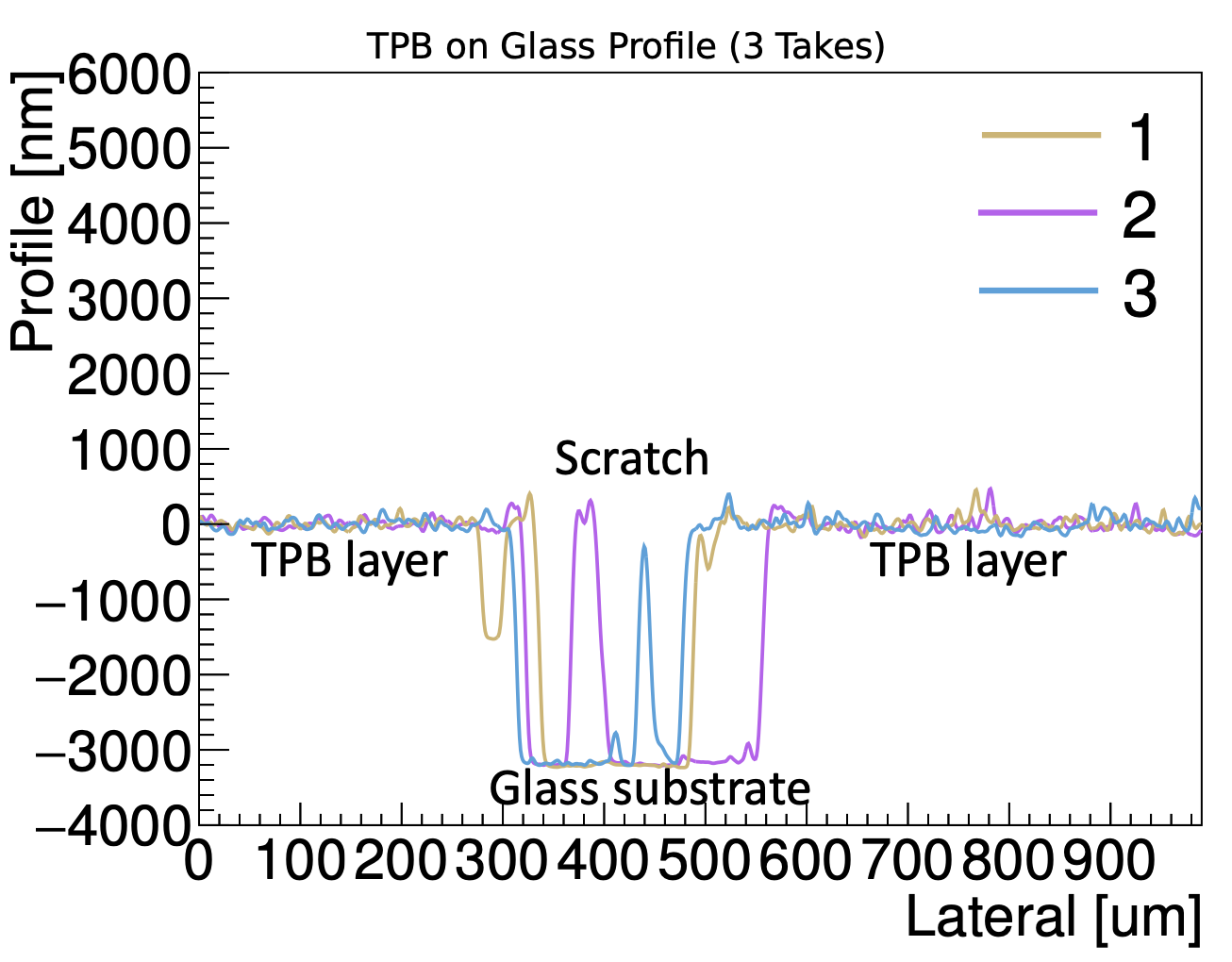}
	\caption{\label{fig:prof_thick}Profilometer profile scans of a TPB layer deposited on top of a glass substrate made by scratching the TPB layer and measuring the depth of the gap created by the scratch.}
\end{figure}

As an independent validation, a profilometer was used to measure the thickness of the TPB layer deposited on a glass substrate. A small scratch was made on the TPB coating to provide a reference edge for the profilometer’s scan. This scratch enabled measurement of the differential height between the coating and the bare glass substrate, as shown in figure~\ref{fig:prof_thick}, which presents three distinct scan profiles. The coating surface and substrate were both set to a baseline of 0~$\upmu$m, while the scratch depth was measured to be approximately 3.1~$\upmu$m, corresponding to the TPB layer thickness. The irregular bumps observed in the scan profiles arise from clumps of TPB formed during scratching. This behavior is typical of organic molecules like TPB, which tend to adhere to surfaces and form uneven deposits when physically disturbed. As a cross-check between the calibration runs and the profilometer measurement, the average measured surface density of 325~$\upmu$g/cm$^{2}$, together with the TPB density of 1.16~g/cm$^3$~\cite{Pollmann:2010gs}, can be used to estimate the layer thickness. The thickness is calculated to be 2.8~$\upmu$m, which is comparable with the independent profilometer measurements of approximately 3.0~$\upmu$m.

\section{Assembly and installation of light enhancers in SBND}

\begin{figure}[H]
	\centering
	\includegraphics[width=0.8\textwidth]{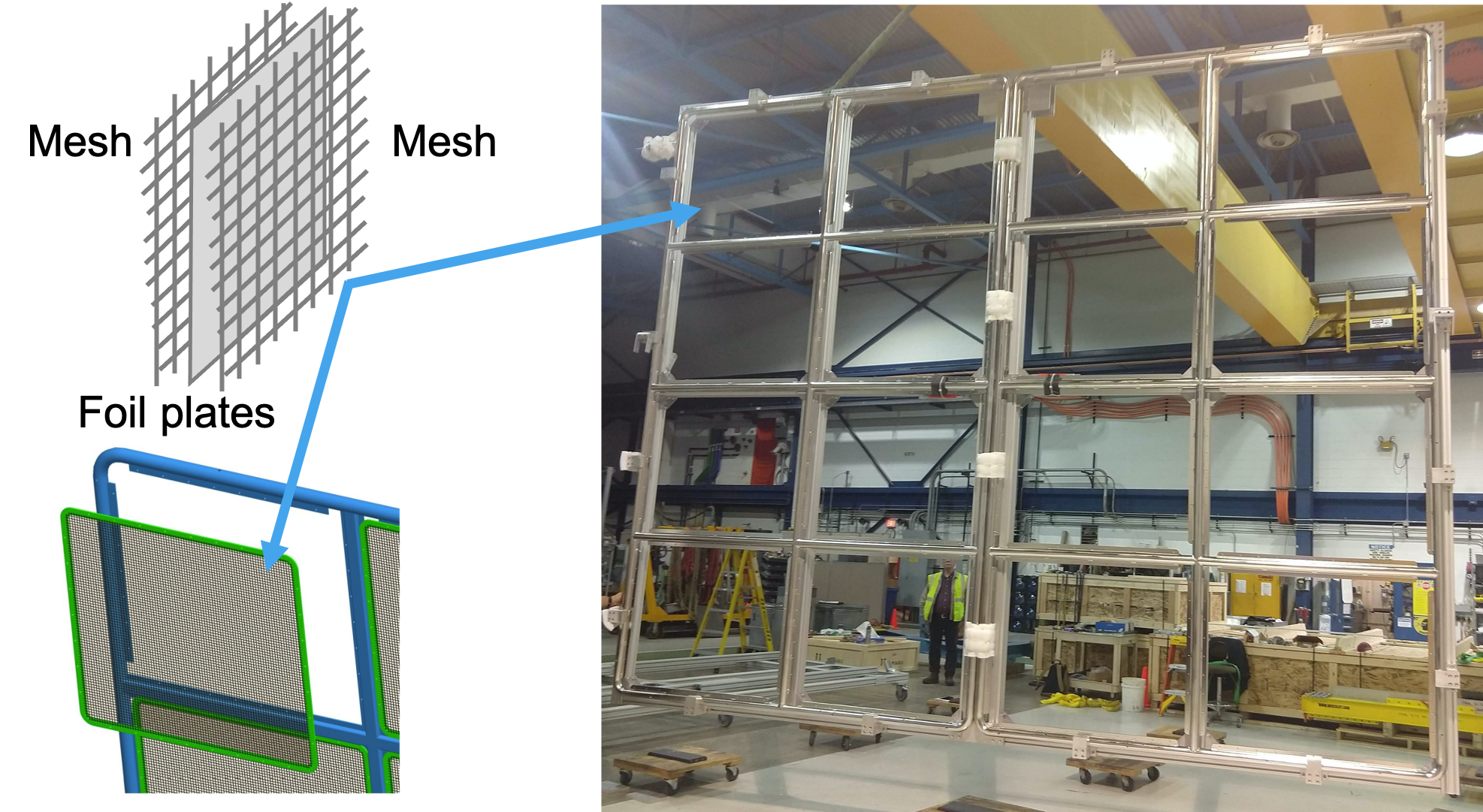}
	\caption{\label{fig:cpa_foil_sandwich}\textit{Left:} Schematics of the mesh-plates-mesh sandwich inside a CPA sub-frame that was mounted into the CPA. \textit{Right:} The CPA frame before inserts were installed.}
\end{figure}

All coated plates were shipped to Fermilab, where they were assembled into 16 mesh assemblies before installation on the cathode frame. As mentioned, the cathode operates at high voltage to establish the TPC drift electric field, and the reflector plates are non-conductive. To prevent disruption of the electrical field from the plates, they were sandwiched between two layers of thin stainless steel mesh during cryofitting of the assembly. The schematics of the mesh-plates-mesh sandwich and the cathode, along with a photograph of the cathode frame, are shown in figure~\ref{fig:cpa_foil_sandwich}. This mesh-plates-mesh sandwich was secured onto the lip of the stainless steel sub-frames, which were then mounted onto the cathode.

\subsection{Assembly}

\begin{figure}[H]
	\centering
	\includegraphics[width=0.45\textwidth]{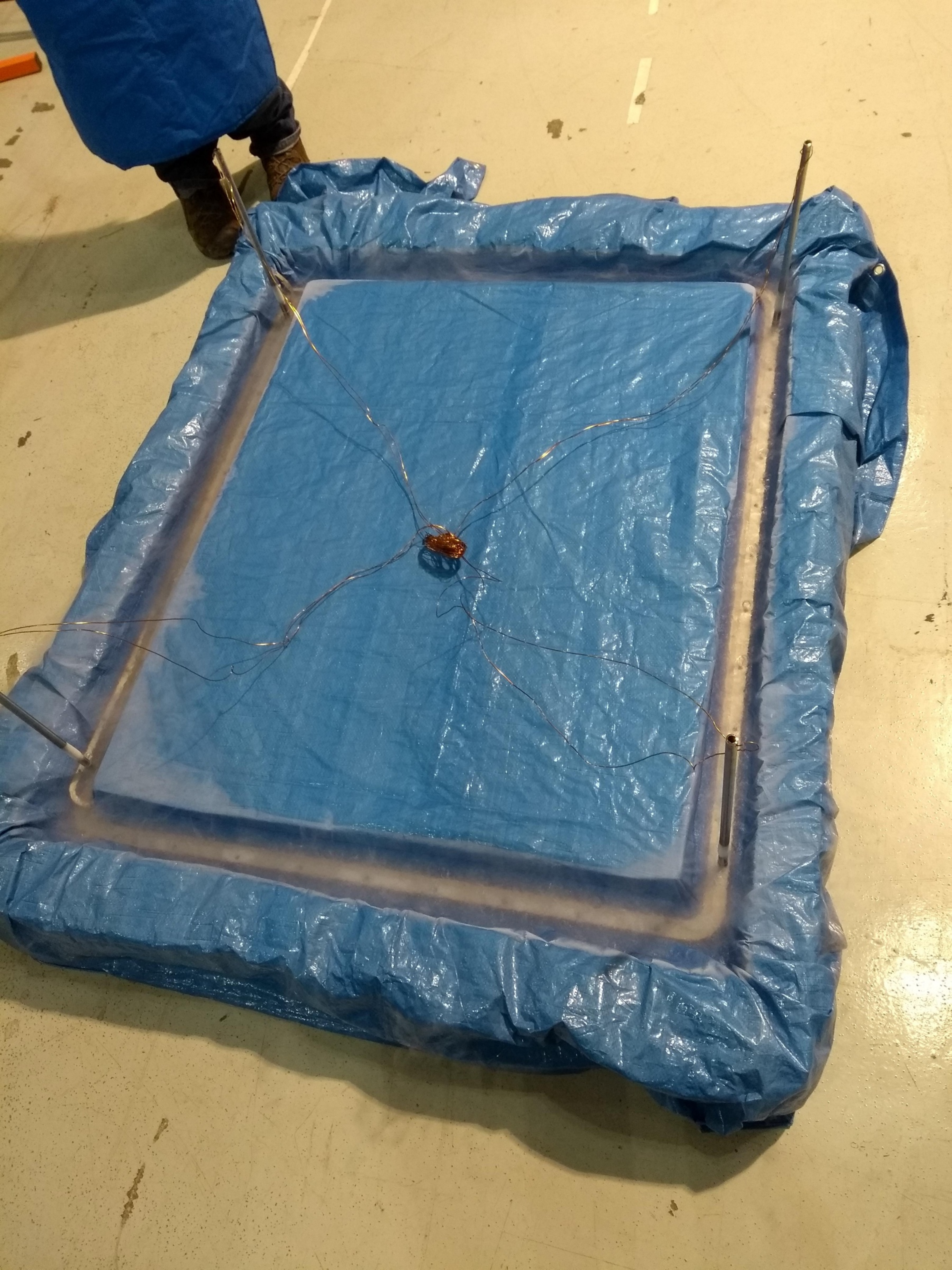}
	\caption{\label{fig:tensioning_2}The sub-frame and its components being cooled down in the liquid nitrogen bath before the assembly.}
\end{figure}

\begin{figure}[H]
	\centering
	\includegraphics[width=0.62\textwidth]{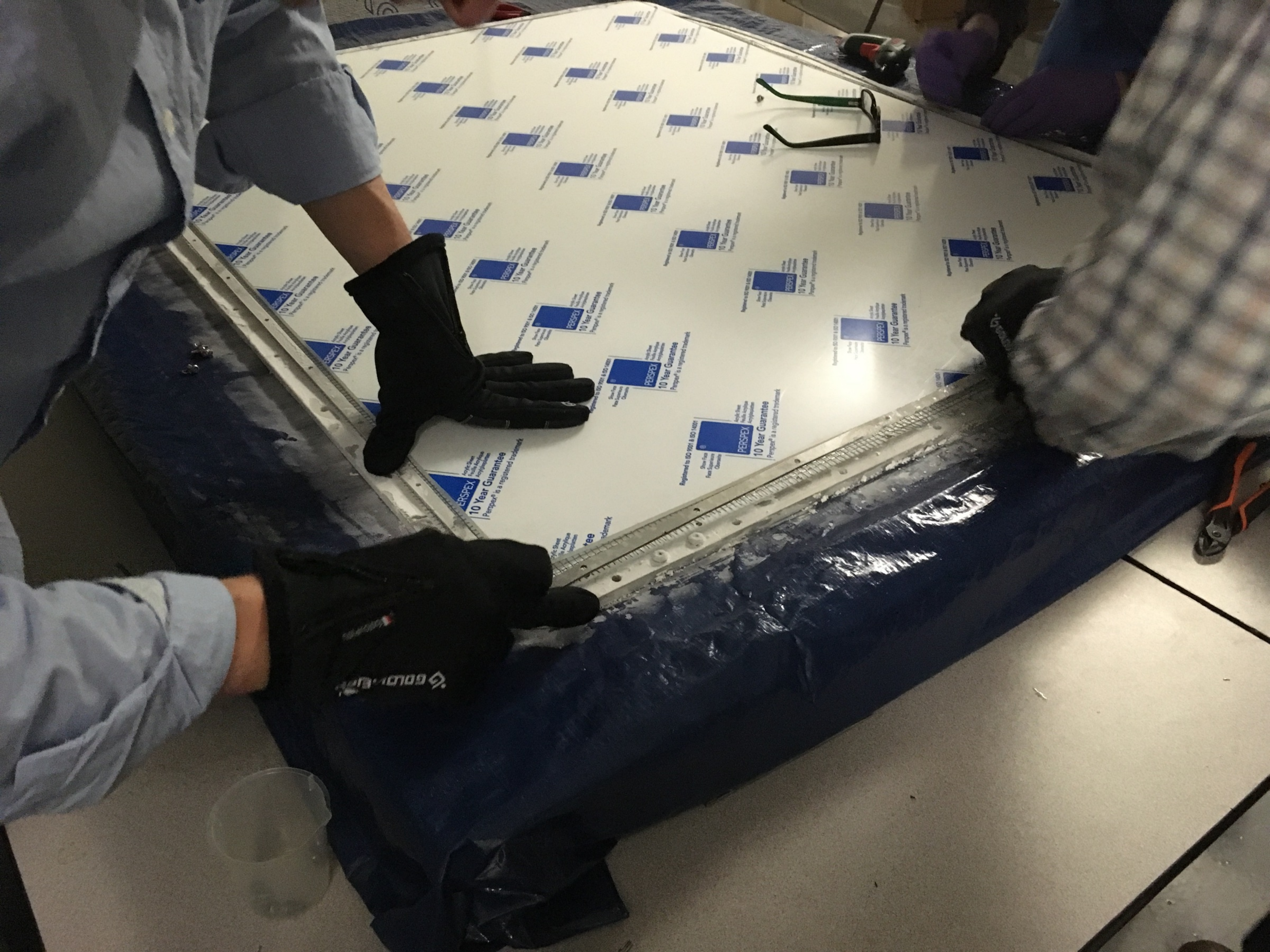}
	\caption{\label{fig:tensioning_5}The process of placing and screwing down the L-shaped strips after sandwiching the plates between two meshes. A second sheet of plastic was placed on top to push down on the mesh and prevent it from moving during the tensioning.}
\end{figure}

\begin{figure}[H]
	\centering
	\includegraphics[width=0.62\textwidth]{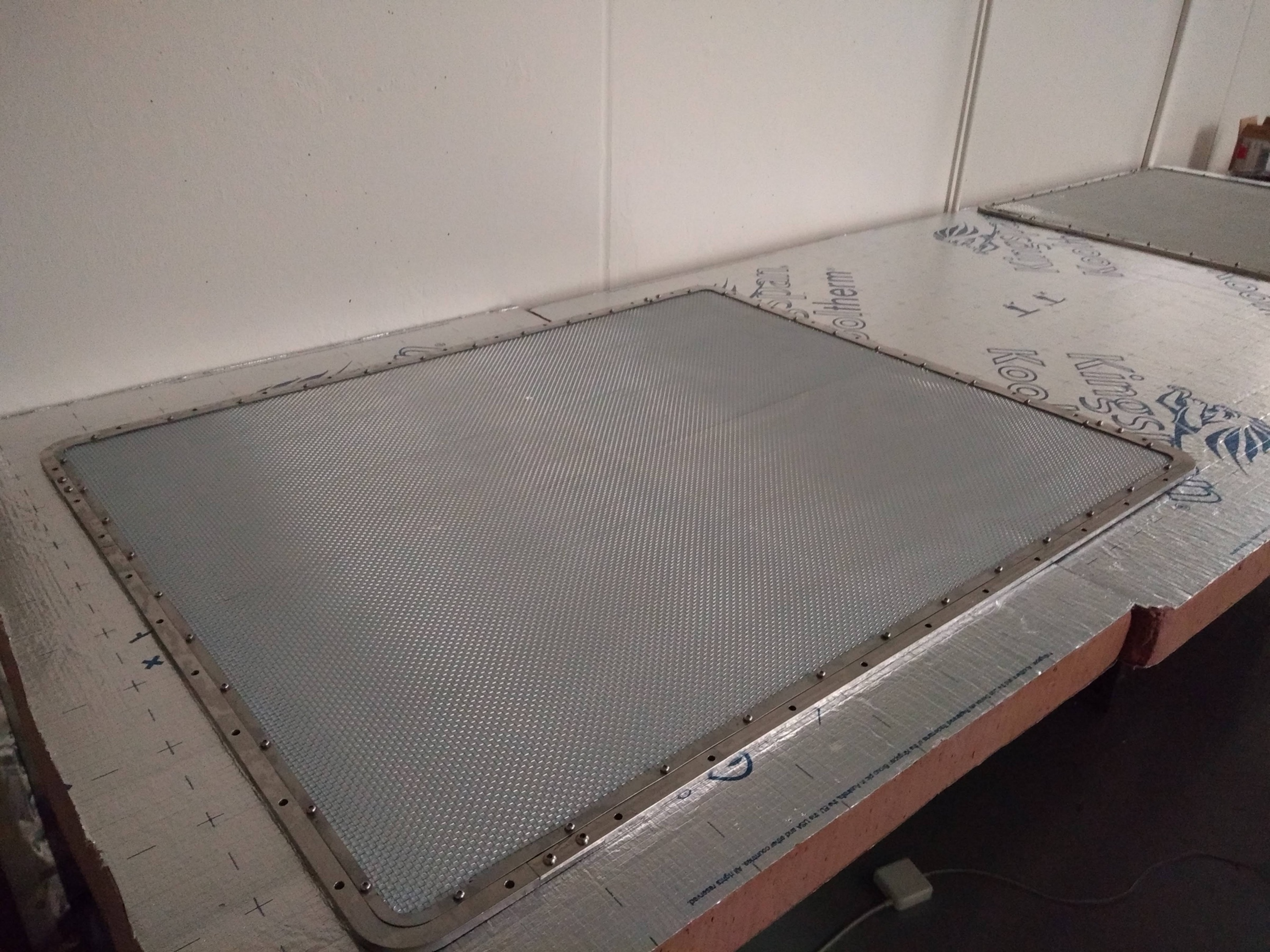}
	\caption{\label{fig:tensioning_6}A mesh-plates-mesh assembly after all four L-shaped washers were placed; the sub-frame was left to dry and warm up before being placed in a light-tight environment.}
\end{figure}

The assembly process involved cooling the empty sub-frame and four long custom-made L-shaped washers that encircled its perimeter with liquid nitrogen (LN$_{2}$) to tension the mesh sandwich onto the sub-frame. This cryofitting method is necessary to keep the cathode frame as flat as possible to maintain the uniformity of the electric field. A thermal bath consisting of two 2-inch-thick Kingspan Koolterm insulation sheets~\cite{kingspan} glued together with cryogenic epoxy (3M Scotch-Weld Epoxy Adhesive~\cite{epoxy}) was fabricated to form a trench, thereby reducing the required quantity of liquid nitrogen. Silicone caulk was used to seal the gaps to prevent spillage, and a polyethylene plastic tarp was placed to cover the entire bath. The sub-frame and washers sat in the trench to cool down safely. Once cooled, the components for a single window were moved to an installation table, where the first mesh was carefully placed inside the sub-frame and aligned with the screw holes. Next, four coated reflector plates were positioned on top, with their screw holes aligned to the sub-frame’s holes. The second mesh was then placed over the plates. To minimize friction and mesh movement and maintain a flat surface, large plastic sheets were placed beneath the bottom mesh and on top of the upper mesh. Each sub-frame’s four long washers are then screwed down using Torx screws around the edges of the mesh sandwich. This step was performed as swiftly as possible to achieve optimal tensioning before the components warmed up. After assembly, the completed sandwich was set aside to fully warm up and dry next to a dehumidifier, which removed residual moisture. Before being stored in light-tight boxes until installation in the detector, each of the screws was verified to have properly latched to mitigate them from loosening due to vibration during the move or during the cool-down of the detector. This process was repeated until all 16 assemblies were completed. Figures~\ref{fig:tensioning_2} through \ref{fig:tensioning_6} illustrate the various stages and components of the tensioning procedure.

\subsection{Installation}

\begin{figure}
	\centering
        \includegraphics[width=0.196\textwidth, angle=0]{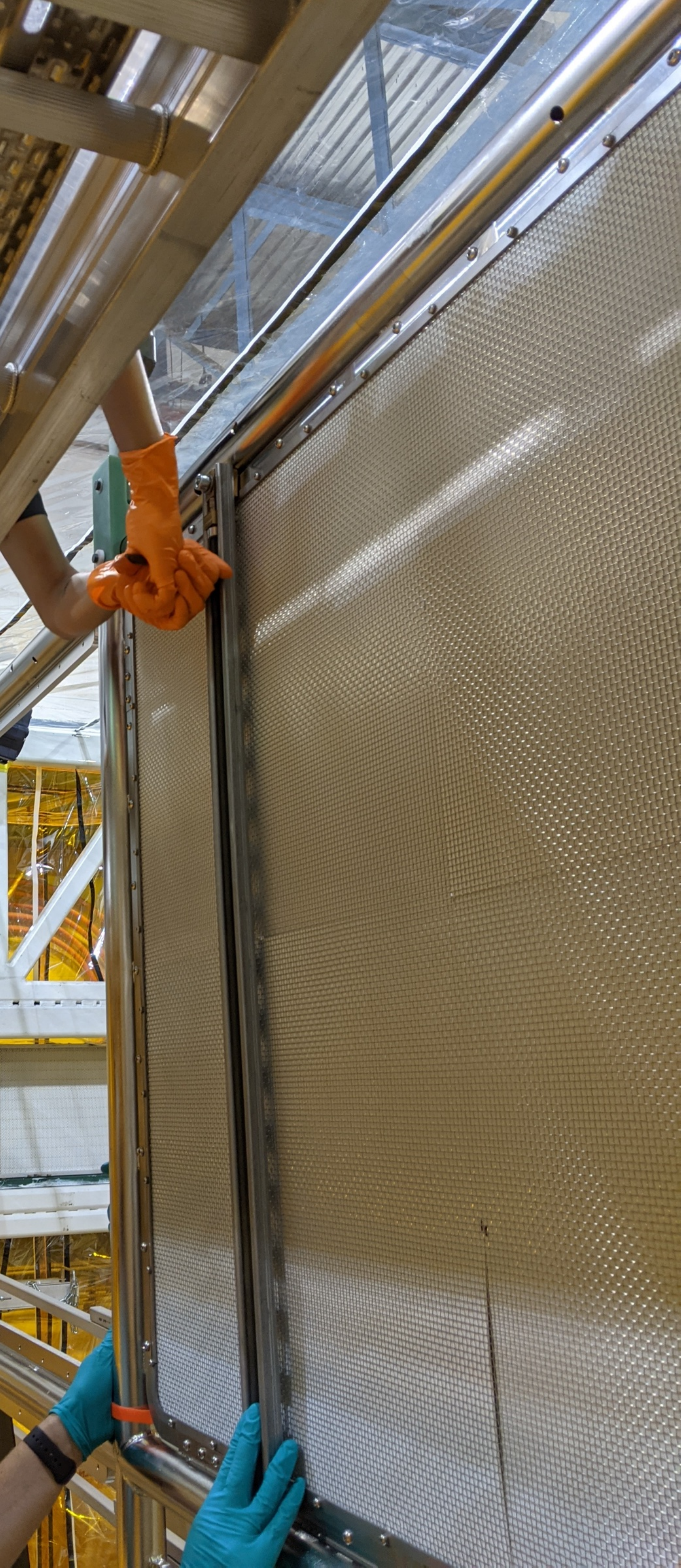}
        \includegraphics[width=0.35\textwidth, angle=0]{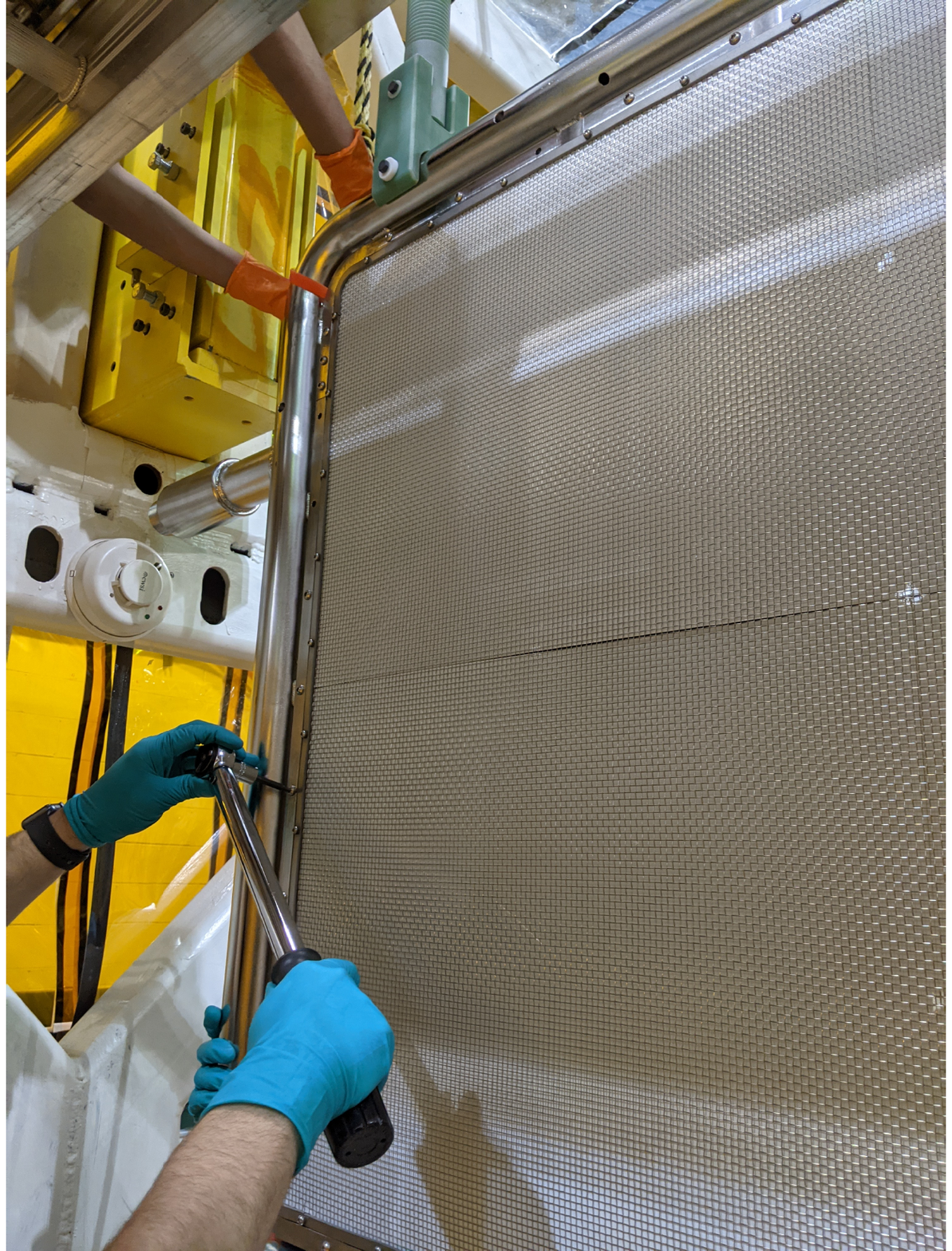}
        \includegraphics[width=0.35\textwidth, angle=0]{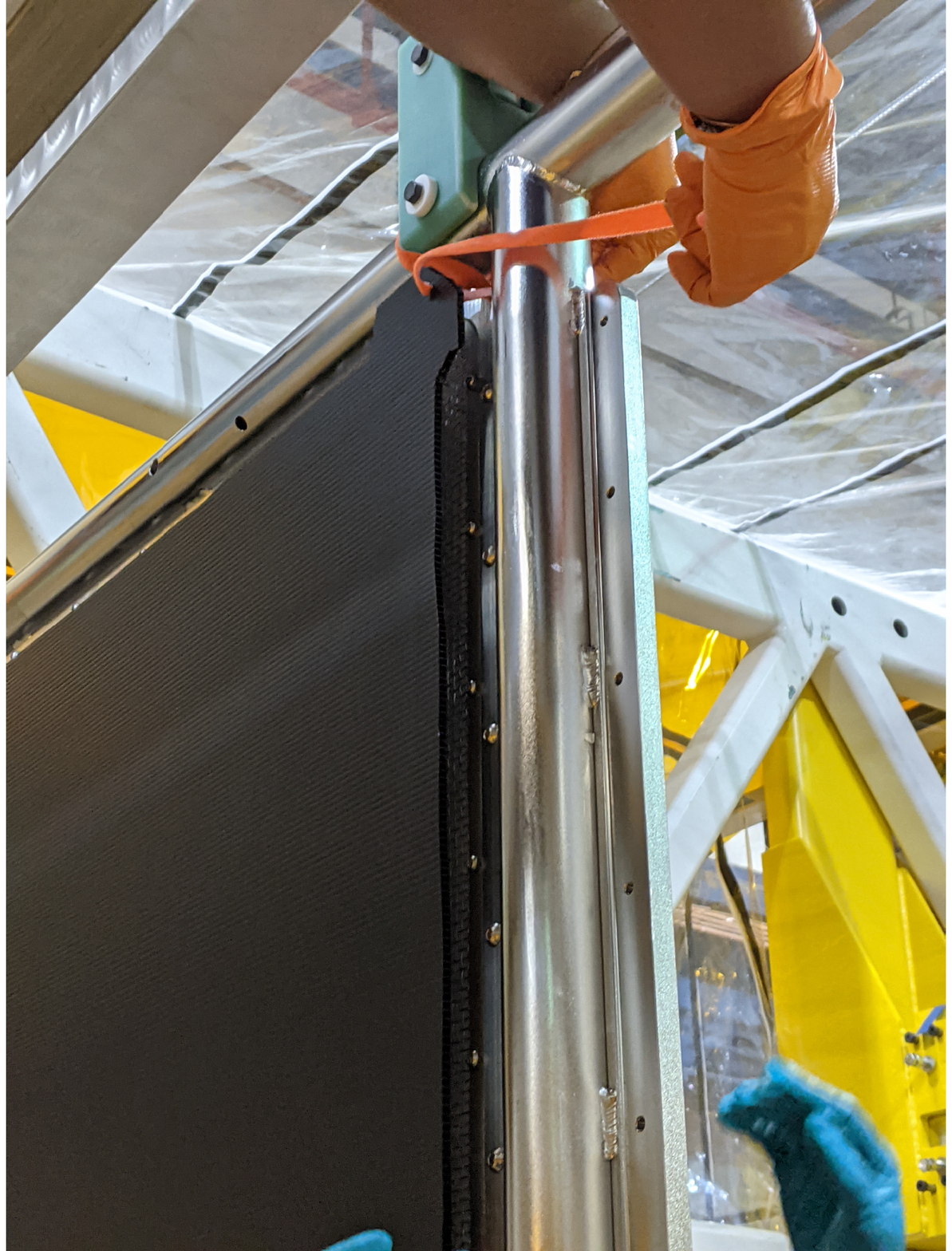}

	\caption{\label{fig:install}\textit{Left:} The stretching tool is being used to align the assembly holes to the ones on the cathode. \textit{center:} A mesh assembly being mounted onto the cathode window and torqued. \textit{Right:} The installation of an opaque cover after the mesh assembly was secured in the window.}
\end{figure}

The mesh assemblies were stored in a dark, dry environment until the final cathode installation. The cathode frame was machined so that the assemblies could be installed from one side and rest on a wide lip machined into the cathode, forming a flat surface level with the stainless steel tubes of the frame. The installation required two individuals on each side of the cathode to place and hold them, working from adjustable-height scaffolding raised to 4~m above the ground due to limited access inside the detector and the height of the assemblies being installed. The assemblies were installed in situ on the cathode while the remaining detector components were being assembled. Each mesh assembly was carefully brought into the detector’s controlled, UV- and blue-light-protected installation area; details about the protection from ambient lighting are presented in section~\ref{sec:protection}. The assemblies were aligned with the cathode windows and secured using bolts inserted into each of the 40 threaded holes in the cathode frame. All bolts were torqued to 8~Nm once fully anchored, and the frame assemblies were secured. To align the frame assemblies with the threaded holes in the cathode, a stretching tool was fabricated to mitigate misalignment. Both ends of the tool were inserted into the opposite side holes and stretched so that the neighboring holes could be aligned and the bolt inserted. Once fully attached, opaque corrugated plastic covers were attached to both sides of each frame to protect the coated surfaces from damage and reduce light exposure during the remainder of the installation. The process was repeated with the next assembly until all 16 were secured in place. Figure~\ref{fig:install} shows photographs of the stretching tool in action to correct misalignment, a mesh assembly being installed and torqued onto the cathode, as well as the placement of an opaque cover to protect the coating.

\section{Protection of wavelength-shifter components from ambient light}{\label{sec:protection}}

More than half of all elements of the PDS of SBND are coated with wavelength-shifting thin films that are sensitive to prolonged exposure to ambient light. Previous studies have shown that TPB coatings degrade more rapidly when exposed to ambient light compared to storage in dark conditions over the same period of time, with the degradation becoming particularly pronounced after approximately 100 days of near-continuous ambient light exposure~\cite{Acciarri:2013rra}. Therefore, it was important to protect the wavelength-shifting coatings to maintain their performance and prevent irreversible damage and loss of conversion efficiency due to the conversion of TPB to benzophenone~\cite{Jones:2012hm}. The PMTs, X-ARAPUCAs, and cathode foils were kept in a dark environment for as long as possible. While the focus of this paper is the reflector plates, the same precautions were also applied to the rest of the PDS.

As mentioned previously, the TPB coating on the reflector plates was applied in a light-tight, low-humidity environment. The plates remained in their nitrogen-filled bags until assembly; the mesh assemblies were then kept in a dark box until installation. During installation on the cathode, the assemblies remained inside the SBND clean tent, with opaque covers protecting the mesh until installation was complete. After installation, the covers were placed back on the meshes. These covers had to be removed several months before the detector was lowered into its cryostat, since access to the cathode became impossible once the detector was fully closed. To account for this, a blue-light-filtering plastic was added as an outer layer on the UV-filtering clean tent to protect the PDS coatings from prolonged exposure to damaging wavelengths.

\begin{figure}[H]
	\centering
	\includegraphics[width=0.9\textwidth]{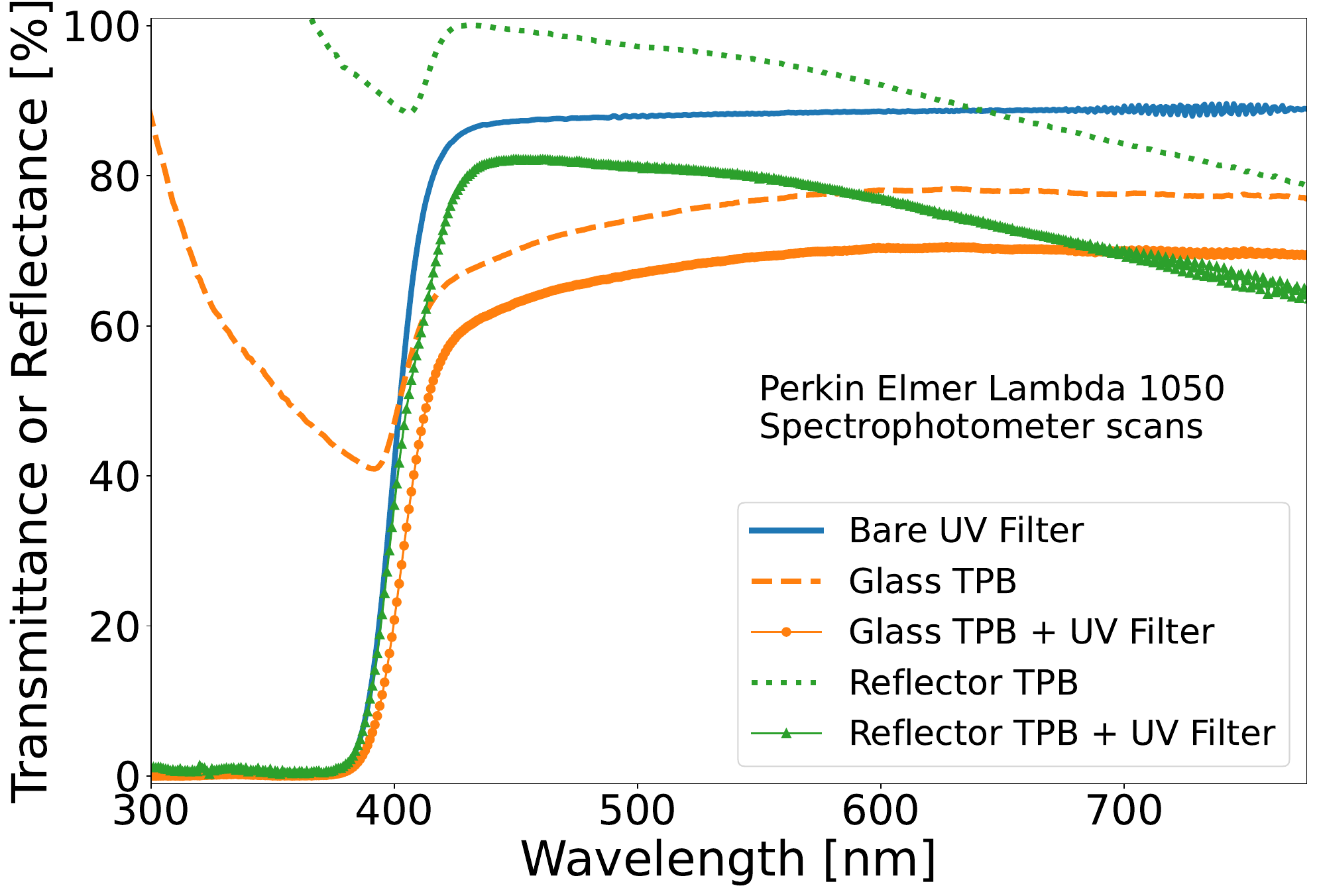}
	\caption{\label{fig:spectrophotometer_results}Spectrum of the transmittance of the different substrates with and without the UV filter plastic. In the case of the reflector substrate, the reflectance is shown instead.}
\end{figure}

The initial UV-only shielding plastic was tested with a PerkinElmer Lambda 1050 Spectrophotometer~\cite{perkin_elmer_lambda}. This Spectrophotometer measured the transmittance and reflectance of different sample configurations used in a selected wavelength range using the integrated sphere setup~\cite{perkin_elmer_sphere}. Figure~\ref{fig:spectrophotometer_results} shows the transmittance (reflectance) of the bare UV plastic filter in blue and TPB coatings on the glass or reflector substrate with and without the filter in orange or green, respectively. As expected, the UV filter fully blocks wavelengths below 380~nm but has minimal effect in the violet-blue region, where TPB can absorb light and lose efficacy during long exposures. Similar to the UV filter, glass is excellent at shielding against wavelengths $<400$~nm, but adding TPB on top of the glass substrate allows light in that region to be recovered, increasing transmittance. In the case of the reflector, the reflectance is optimized in the visible range and, like the glass, also recovers in the $<400$~nm region due to the wavelength-shifting of the TPB.

\begin{figure}[H]
	\centering
	\includegraphics[width=0.9\textwidth]{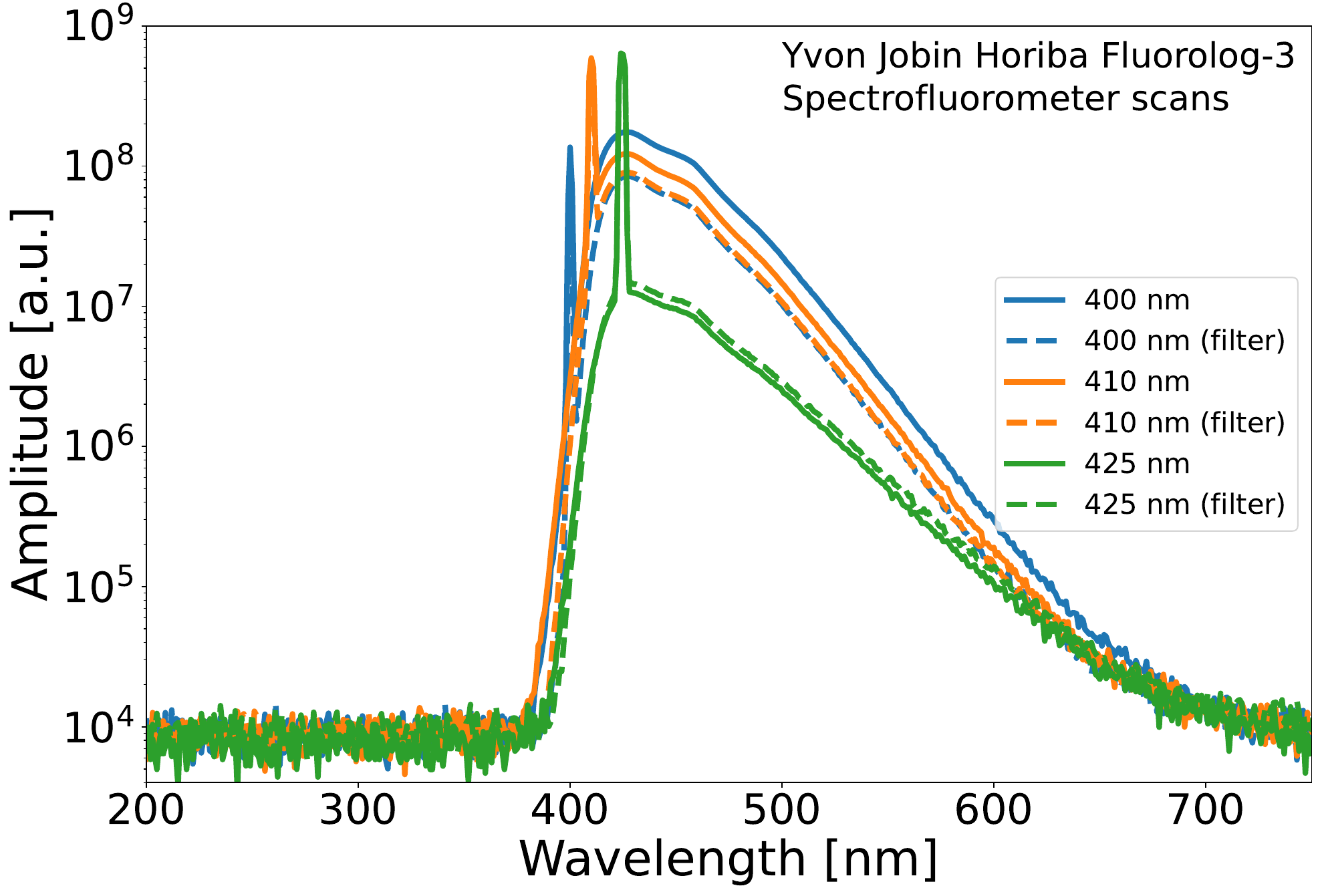}
	\caption{\label{fig:fluorolog_results}Emission spectrum of TPB on glass for different excitation wavelengths from 400~nm to 425~nm and comparison with and without the UV filter. The spikes at 400~nm, 410~nm, and 425~nm are artefacts caused by the excitation wavelength being measured with the emission monochromator. The UV filter has minimal effect on preventing TPB fluorescence in the blue-light range.}
\end{figure}

As a verification that wavelengths passing through the UV filter could still be absorbed and re-emitted by TPB, TPB on glass with and without the filters was tested using a Yvon Jobin Horiba Fluorolog-3 Spectrofluorometer~\cite{fluorolog_spec}. This setup allowed the measurement of the emission spectrum at different wavelengths for specific excitation wavelengths. Figure~\ref{fig:fluorolog_results} shows the results of these scans with different excitation wavelengths to determine the intensity of emission from TPB as a function of the excitation wavelength, and the effect of adding the UV blocking filter on top of the TPB layer. The emission amplitude decreases as the wavelength increases, indicating that TPB emission has an upper wavelength limit and that the UV filter does not shield the TPB from wavelengths in this range, as shown by the spectrophotometer scans.

\begin{figure}[H]
	\centering
	\includegraphics[width=0.7\textwidth]{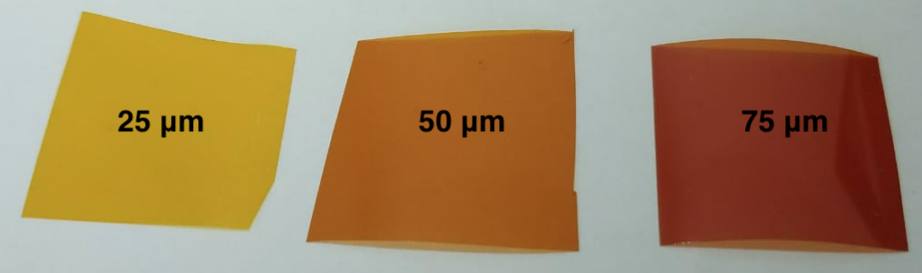}\\
	\centering
	\includegraphics[width=0.8\textwidth]{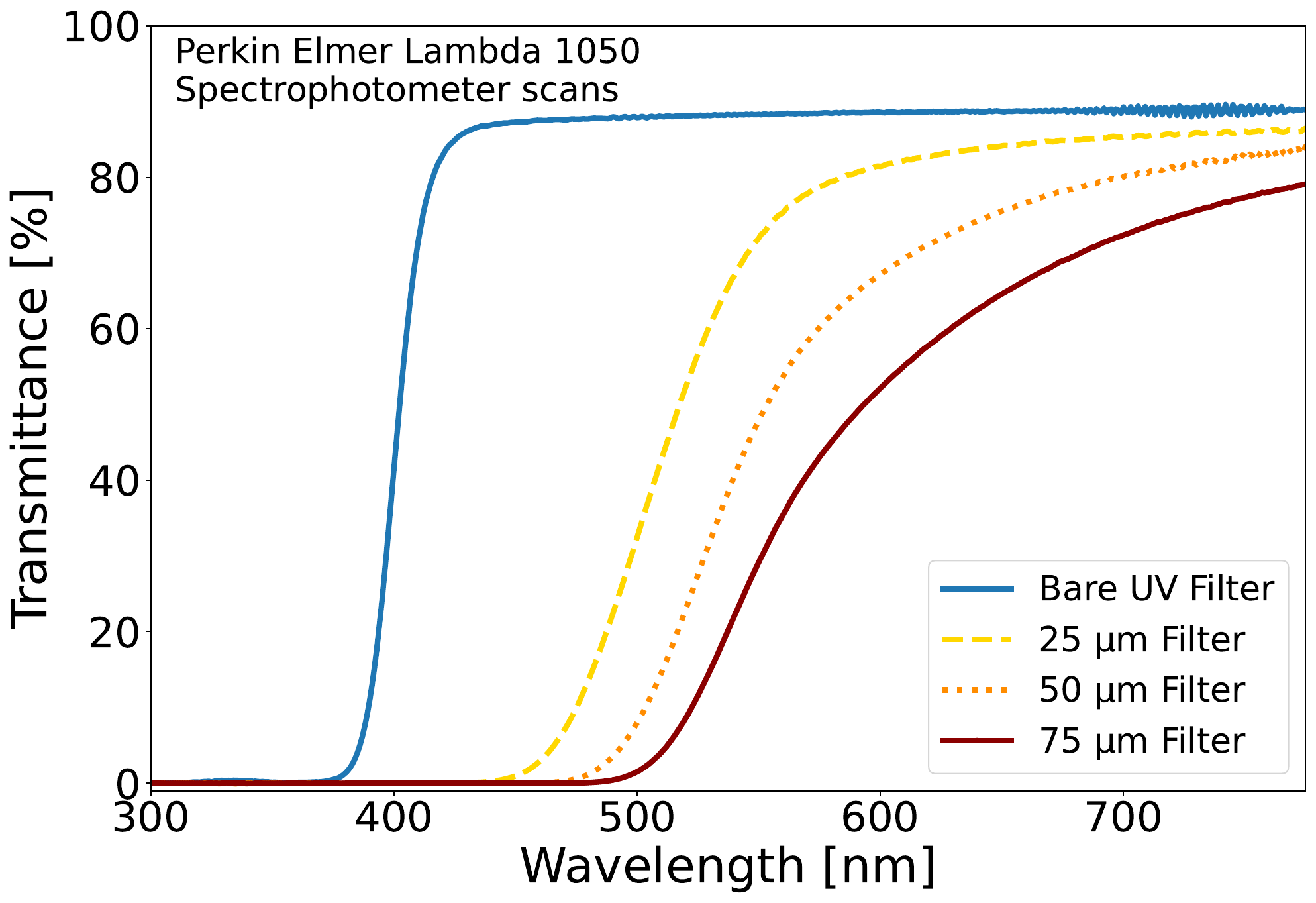}
	\caption{\label{fig:blue_filter} Different Kapton samples selected for the blue light filter of different thicknesses and blocking wavelengths compared to the bare UV filter. The 50~$\upmu$m offers the best cut-off wavelengths after 450~nm while also not creating a too dark work environment.}
\end{figure}

Following these findings, it was determined that an additional layer of protection was necessary due to extended exposure to ambient light, which may weaken bonds and reduce conversion efficiency by the time the TPC is placed into the cryostat. Three UV and blue-light plastic filters were tested again using the Spectrophotometer to measure their transmittance, as shown in figure~\ref{fig:blue_filter}, which indicates that all three filters would prevent light below 450~nm from entering the installation area. The 50~$\upmu$m sample was selected because it shields well below 450~nm without making the clean tent too dark to work in.

\begin{figure}[H]
	\centering
	\includegraphics[width=0.85\textwidth]{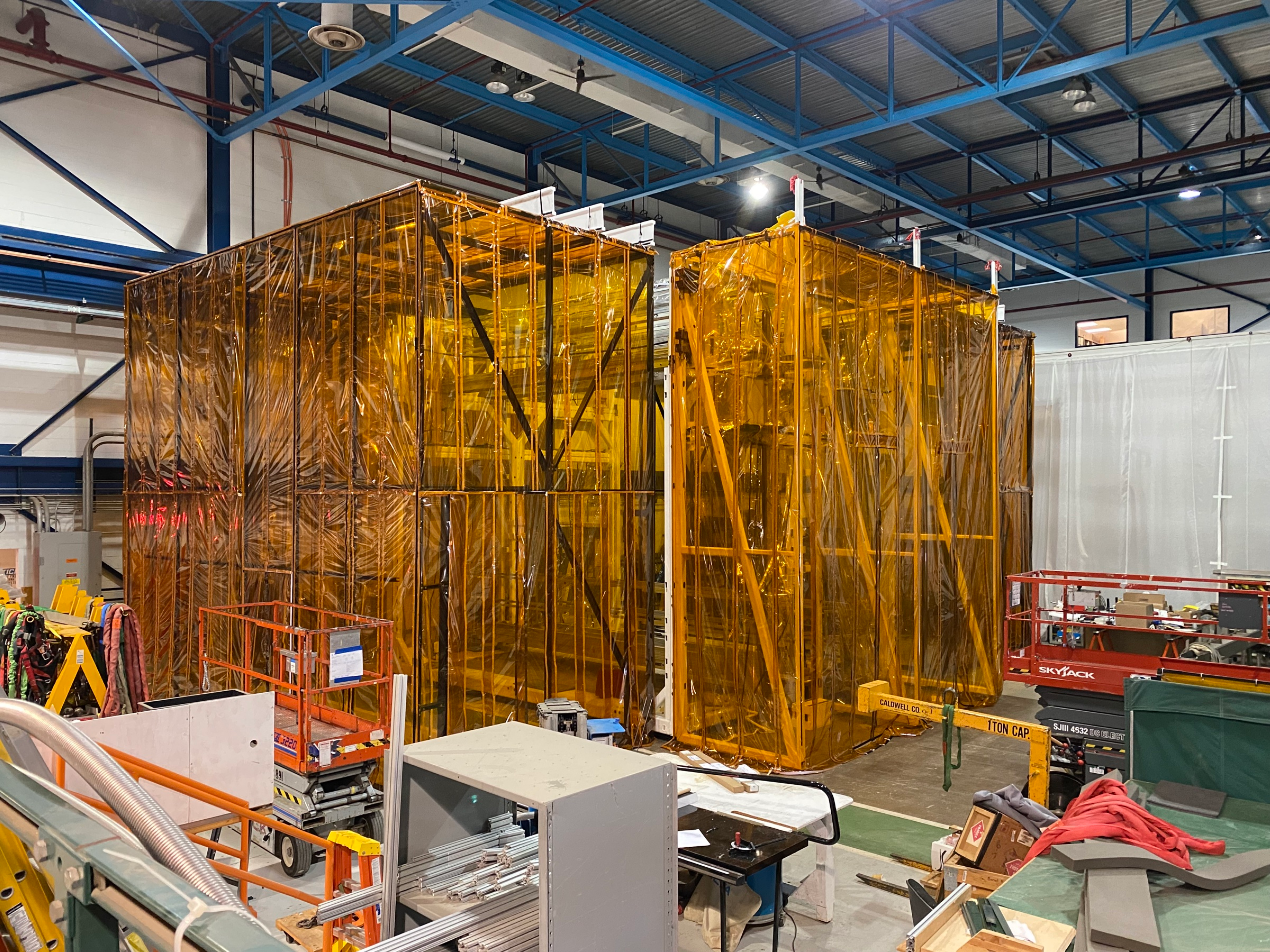}	\caption{\label{fig:clean_tent}The SBND installation clean tent. The blue light filter covered the outer layer in a yellow-orange color, preventing ambient blue light from entering the area.}
\end{figure}

A photograph of the SBND clean tent is shown in figure \ref{fig:clean_tent}. The clear UV filter plastic is placed in the central region, while the blue light filter appears yellow-orange. This dual-layered tent was kept in place throughout the detector’s assembly and installation until it was lowered into the cryostat, thereby minimising exposure to ambient light.

\section{Summary}

This article describes the production, assembly, and installation of the wavelength-shifting reflective plates in SBND, which serve as a passive light collection enhancing system. This endeavor represents the largest area of a neutrino detector covered with TPB to date. Just over 80 plates were produced during three evaporation campaigns at the University of Manchester and UNICAMP. Following this, the plates were assembled and installed onto the TPC cathode. The system has already yielded preliminary results, showing both the expected increase in light yield and improved uniformity across the drift direction~\cite{sbnd_detector_paper}. Both the uncoated PMTs and uncoated X-ARAPUCAs have detected signals originating from the reflective surfaces. This effort has been successful and is expected to meet the performance goals outlined in ref.~\cite{SBND:2024vgn}. The methods and results presented in this article provide an important set of techniques and lessons learned for future light-enhancement efforts in liquid noble element detectors. Scaling evaporative TPB coatings to the much larger surface areas required for experiments such as DUNE presents additional challenges, including ensuring mechanical robustness during handling and installation and maintaining production throughput and quality control at industrial scales. While vacuum evaporation onto large surfaces is non-trivial, the experience gained in SBND demonstrates that controlled, uniform coatings are achievable and provides a foundation for further R\&D toward next-generation detectors.

\acknowledgments
This document was prepared by the SBND Collaboration using the resources of the Fermi National Accelerator Laboratory (Fermilab), a U.S. Department of Energy, Office of Science, Office of High Energy Physics HEP User Facility. Fermilab is managed by FermiForward Discovery Group, LLC, acting under Contract No. 89243024CSC000002. The SBND Collaboration acknowledges the generous support of the following organizations: the U.S. Department of Energy, Office of Science, Office of High Energy Physics; the U.S. National Science Foundation; Los Alamos National Laboratory for LDRD funding, the Science and Technology Facilities Council (STFC), part of United Kingdom Research and Innovation (UKRI), the UKRI Future Leaders Fellowship (grant number MR/V022407/1), and The Royal Society; the Swiss National Science Foundation; the Spanish Ministerio de Ciencia, Innovacíon y Universidades (MICIU/ AEI/ 10.13039/ 501100011033) under grants No PRE2019-090468, CNS2022-136022, RYC2022-036471-I, PID2023-147949NB-C51 \& C53 and Comunidad de Madrid (PEJ-2023-AI/COM-28399); the European Union’s Horizon 2020 research and innovation program under GA no 101004761 and the Marie Sklodowska-Curie grant agreements No 822185, 101081478, and 101003460; the São Paulo Research Foundation 1098 (FAPESP), the National Council of Scientific and Technological Development (CNPq) and Ministry of Science, Technology \& Innovations-MCTI of Brazil; the Minas Gerais research foundation (FAPEMIG), grants APQ-00544-23 and APQ-01249-24; the Anusandhan National Research Foundation (ANRF, India) under the Ramanujan Fellowship (Grant No. RJF/2025/000203). An award of computer time was provided by the ASCR Leadership Computing Challenge (ALCC) program. This research used resources of the Argonne Leadership Computing Facility, which is a U.S. Department of Energy Office of Science User Facility operated under contract DE-AC02-06CH11357.

\bibliographystyle{JHEP}
\bibliography{main}

\end{document}